\documentclass[]{aa} 
\usepackage{graphicx}
\usepackage{txfonts}
\usepackage{color}
\usepackage{float}
\usepackage[normalem]{ulem}
\usepackage{xcolor}
\usepackage[fleqn]{amsmath}
\usepackage{multicol}
\usepackage{multirow}
\usepackage{hyperref}
\hypersetup{colorlinks = true, linkcolor = {blue}, citecolor = {blue}, urlcolor= {blue}}
\newcommand{\aas}{Bulletin of the American Astronomical Society}
\newcommand{\psj}{The Planetary Science Journal}

\makeatletter
\@ifundefined{linenumbers}{}{%
  \renewcommand{\linenumbers}{}
  
  \renewcommand{\modulolinenumbers}[1]{}
  
}
\makeatother

\begin{document} 

\title{Life in the dark}

\subtitle{Potential urability of moons of rogue planets}

\author{
Viktória Fröhlich
    \inst{1,2,3}\thanks{\href{mailto:frohlich.viktoria@csfk.org}{frohlich.viktoria@csfk.org}}\orcid{0000-0003-3780-7185}
\and
Zsolt Regály
    \inst{1,2}\orcid{(0000-0001-5573-8190)}
}

\institute
{HUN-REN Konkoly Observatory, Research Centre for Astronomy and Earth Science, Konkoly-Thege Mikl\'os 15-17, 1121, Budapest, Hungary
\and
CSFK, MTA Centre of Excellence, Budapest, Konkoly Thege Miklós 15-17, H-1121, Budapest, Hungary
\and
ELTE E\"otv\"os Lor\'and University, Institute of Physics and Astronomy, P\'azm\'any P\'eter s\'et\'any 1/A, 1171 Budapest, Hungary
}

\date{Received June 23, 2025; Accepted November 4, 2025}

\abstract
{
Free‑floating (rogue) planets are thought to be numerous in the Galaxy and may retain their moons after ejection from their natal systems. 
If those satellites acquire or preserve orbital eccentricity, tidal dissipation could provide a long‑lasting internal heat source, potentially creating urable environments (capable of enabling abiogenesis) in the absence of stellar radiation.
}
{
We explore (i) whether moons remain dynamically bound to planets expelled by a core‑collapse (Type~II) supernova, (ii) how the explosion reshapes their orbits, and (iii) under which circumstances tidal heating can sustain urable subsurface oceans that meet the minimal conditions for life to originate.
}
{
We carried out 4,412 two‑dimensional N‑body simulations with an 8th‑order Runge–Kutta scheme, modelling homologous stellar mass loss for progenitors of $10~M_\odot$. 
Post‑explosion orbital elements of single moons and resonant moon systems were analysed, and tidal heating power was estimated with a constant phase‑lag prescription for several tidal dissipation functions and moon densities.
}
{
All simulated moons survive the supernova and remain bound to their planets. 
The explosion excites moon eccentricities up to $\simeq 7 \times 10^{-4}$ and $\simeq 3 \times 10^{-3}$ for single moons of planets with circular and eccentric orbits, respectively.
For resonant pairs, an eccentricity of $2\lesssim \times10^{-2}$ is preserved.
The semi‑major axis of the moons changes by $\lesssim 0.2~\%$. 
For $12-15~\%$ of cases -- preferentially moons at $a\leq15~R_\mathrm{p}$ and with $e\geq~10^{-3}$ -- the specific tidal heating power lies between 0.1 and 10 times what is estimated on Europa or Enceladus, sufficient to maintain liquid oceans beneath an ice crust. 
Eccentricity damping timescales exceed the age of the Solar System for $a\geq10~R_{\mathrm{p}}$, implying billions of years of continuous heating on the moons.
}
{
Moons of rogue planets ejected via Type~II supernova explosions are both dynamically stable and, in a significant minority of configurations, tidally active enough to host long‑lived subsurface oceans. 
Such worlds represent promising targets for future searches for extraterrestrial life.
}

\keywords{
methods: numerical -- planets and satellites: dynamical evolution and stability -- stars: evolution -- supernovae: general
}

\maketitle

\section{Introduction}
\label{sec:intro}

The number of discovered rogue planets (i.e., planets without a host star, also called nomadic, unbound, orphan, wandering, starless, sunless or free-floating planets) has now reached several hundred (see, e.g. \citealp{lucasroche00, sumietal11, liuetal13, bennettetal14, luhman14, henderson16, miretroigetal21, pearson-mccaughrean-23}).
What is more, theories suggest that there could be as many as two Jupiter-sized rogue planets for every star in the Galaxy \citep{sumietal11, clantongaudi16}.

Several theories can explain the ejection of planets from their host star system.
The most common hypotheses are the interaction of planets with each other \citep{levisonetal98, fordrasio08, guillochanetal11} or with stars \citep{holmanwiegert99, musielaketal05, doolinblundell11, malmbergetal11, verasraymond12, kaibetal13}, leading to early dynamical instabilities.
The other effective channel of rogue planet formation is the post–main–sequence mass loss of the host star, where planetary orbits widen and their eccentricity is pumped above unity, leaving their host star in all but the most extreme cases \citep{verasetal11, verastout12, regalyetal22}.
Note, however, that it cannot be ruled out that rogue planets are also formed during the fragmentation of low-mass gas nebulae \citep{boydwhitworth05}.

\begin{table*}[ht!]
\caption{\centering Parameter space used for modelling the SN II explosion.}
\centering
\begin{tabular}{|c|l|cccc|}
\hline
 &
  Parameter &
  \multicolumn{1}{c|}{$e_{\mathrm{p}}^{(0)} = 0$} &
  \multicolumn{1}{c|}{$e_{\mathrm{p}}^{(0)}\neq 0$} &
  \multicolumn{1}{c|}{$e_{\mathrm{m}}^{(0)}\neq 0$} &
  $n_{\mathrm{m}}^{(0)}:n_{\mathrm{res}}^{(0)}$ \\ \hline \hline
\multirow{3}{*}{\rotatebox{90}{Stars}} &
  $M_{\mathrm{ej}}~(M_\odot)$ &
  \multicolumn{4}{c|}{8} \\ \cline{2-6} 
 &
  $M_{\mathrm{n}}~(M_\odot)$ &
  \multicolumn{4}{c|}{2} \\ \cline{2-6} 
 &
  $v_{\mathrm{max}}~({\mathrm{km~s^{-1}}})$ &
  \multicolumn{3}{c|}{1,000; 10,000} &
  10,000 \\ \hline \hline
\multirow{4}{*}{\rotatebox{90}{Planets}} &
  $M_{\mathrm{p}}~(M_{\mathrm{Jup}})$ &
  \multicolumn{4}{c|}{0.01; 0,1; 1; 10} \\ \cline{2-6} 
 &
  $a_{\mathrm{p}}^{(0)}~(\mathrm{CsE})$ &
  \multicolumn{1}{c|}{{[}5.14:200{]}\tablefootmark{*}} &
  \multicolumn{2}{c|}{13.95} &
  {[}5.14:200{]}$^*$ \\ \cline{2-6} 
 &
  $e_{\mathrm{p}}^{(0)}$ &
  \multicolumn{1}{c|}{0} &
  \multicolumn{1}{c|}{0.1; 0.2; 0.4; 0.6; 0.8} &
  \multicolumn{2}{c|}{0} \\ \cline{2-6} 
 &
  $f_{\mathrm{p}}^{(0)}~(^\circ)$ &
  \multicolumn{1}{c|}{0} &
  \multicolumn{1}{c|}{0, 90, 180, 270} &
  \multicolumn{2}{c|}{0} \\ \hline \hline
\multirow{6}{*}{\rotatebox{90}{Moons}} &
  $M_{\mathrm{m}}~(M_{\mathrm{p}})$ &
  \multicolumn{4}{c|}{$2.5 \times 10^{-5}$} \\ \cline{2-6} 
 &
  $a_{\mathrm{m}}^{(0)}~(R_{\mathrm{p}})$ &
  \multicolumn{3}{c|}{5.89; 9.39; 14.97; 26.33} &
  9.39 \\ \cline{2-6} 
 &
  $e_{\mathrm{m}}^{(0)}$ &
  \multicolumn{2}{c|}{0} &
  \multicolumn{1}{c|}{$10^{-4}$; $10^{-3}$; $10^{-2}$; $10^{-1}$} &
  0 \\ \cline{2-6} 
 &
  $f_{\mathrm{m}}^{(0)}~(^\circ)$ &
  \multicolumn{4}{c|}{0; 90; 180; 270} \\ \cline{2-6} 
 &
  $M_{\mathrm{res}}~(M_{\mathrm{m}})$ &
  \multicolumn{3}{c|}{-} &
  0.5; 1; 2 \\ \cline{2-6} 
 &
  Res. &
  \multicolumn{3}{c|}{-} &
  2:1; 3:1; 3:2 \\ \hline
\end{tabular}
\label{tab:params}
\tablefoot{\centering\tablefoottext{*}{Logarithmic sampling of 12 points on the indicated interval.}}
\end{table*}

Even though over 5,900 exoplanets have been discovered\footnote{As of July 2025., see the \href{https://exoplanetarchive.ipac.caltech.edu/}{NASA Exoplanet Archive}}, and the pattern of the Solar System suggests that these objects may be among the most common in the Galaxy, the existence of the exomoons that putatively orbit them remains a mystery. 
Over the past twenty-five years, dozens of studies have been directed at detecting exomoons, based primarily on direct imaging, joint planet-moon transit light curve modelling, and the modelling and detection of microlensing effects caused by the two bodies (for a detailed overview, see \citealp{heller18}).
The photometric sensitivity of the Kepler space telescope allows for the detection of moons with masses up to 0.2 Earth masses \citep{kippingetal09}.
Nevertheless, the existence of the moons of K1625b \citep{teacheykipping18} and K1708b \citep{kippingetal22} is a matter of ongoing debate \citep{helleretal19, kreidbergetal19, kalmanetal23}.
Furthermore, the six exomoons detected by \citet{foxwiegert21} have been refuted by multiple authors \citep{kipping20, quarlesetal20}.

If a planet with a moon or moons leaves its system, the moons will most likely continue to orbit the rogue planet.
This is true for moons of Earth-like planets ejected by the gravitational perturbations of gas giants \citep{debessigurdsson07}, and also for moons orbiting gas giants that undergo planet-planet scattering \citep{hongetal18, rabagosteffen18}.
Theoretical considerations suggest that around 10-15\% of rogue planets may have moons whose presence can be detected by direct imaging, microlensing or planetary occultations.

The recently introduced concept of urability \citep{deameretal22} describes the potential of an environment to give rise to life, rather than to sustain it. 
While habitability concerns conditions that allow surface liquid water and the persistence of life, urability addresses the minimal physical, chemical, and energetic requirements for life to originate. 
This distinction is crucial when assessing worlds that are deprived of stellar irradiation. 
In such systems, internal and tidal heat sources may still generate local environments capable of prebiotic activity. 
In this context, we investigate whether moons orbiting rogue planets can maintain subsurface oceans heated by tidal dissipation, and whether they could meet the basic criteria for urability.

In this study, we present a dynamical study to model the birth of rogue planet-moon systems following a Type II supernova (SN~II) explosion.
We study the change in the systems' orbital elements caused by the mass loss of the SN~II numerically with a homologous envelope expansion model.
The paper is structured as follows.
Section~\ref{sec:methods} details the adopted numerical methods.
Section~\ref{sec:results} lists the results of the numerical simulations.
In Sect.~\ref{sec:disc} we discuss our results and give a brief analysis of the tidal heating on rogue moons to see whether they could harbour environments capable of starting life.
We give our conclusions in Sect.~\ref{sec:conclusions}.

\section{Numerical methods}
\label{sec:methods}

In this study, we investigated the orbital dynamics of an exomoon orbiting a planet whose host star explodes as an SN~II.
During the SN~II explosion, the central star undergoes significant mass loss, perturbing the orbits of the planet-moon system.
The stellar mass inside the planet's and the moon's orbits changed according to a homologous expansion model.
The equations of motion for the bodies were solved numerically in two dimensions using an 8th-order Runge--Kutta method.
The numerical methods are presented in detail in Appendix~\ref{sec:apx:homologous}. 

During the simulations, we followed the orbital parameters of the components and considered a planet or moon bound if its eccentricity remained under unity.
Throughout the paper, orbital elements are labelled with a superscript $^{(0)}$ pre-, and a superscript $^{(1)}$ post-explosion (e.g. $e_{\mathrm{m}}^{(0)} \xrightarrow{} e_{\mathrm{m}}^{(1)}$).
We note that in a three-body problem with no mass loss, the eccentricity of the moons oscillates between $10^{-8}$ and $10^{-6}$. 
This is because in a hierarchical three-body problem, circular orbits do not exist.

\begin{figure}[t!]
    \centering
    \includegraphics[width=0.995\columnwidth]{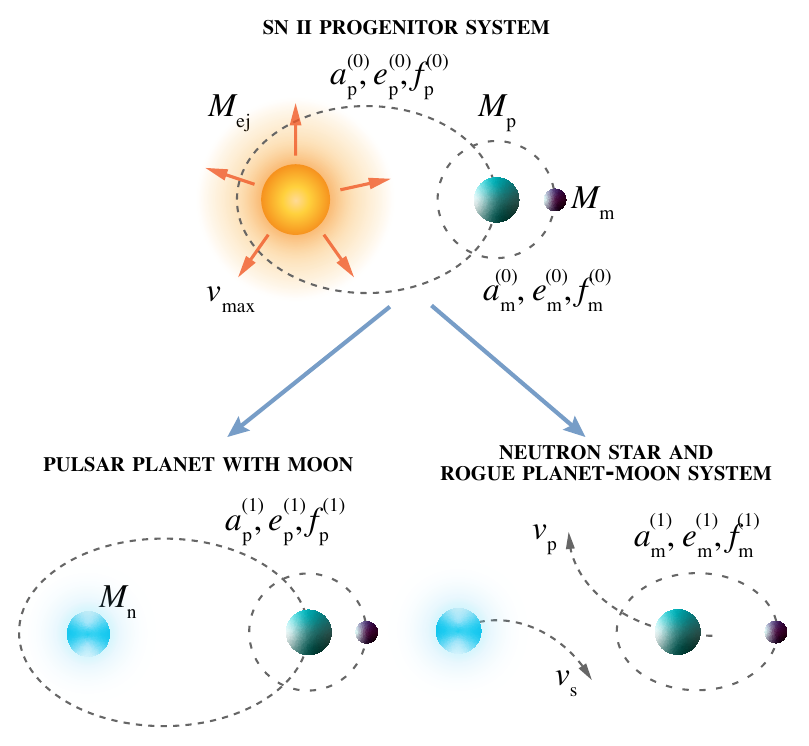}
    \caption{
    The visual representation of the parameters used for modelling the SN~II explosion. 
    The progenitor system (top) can evolve into two distinct end states (bottom): 
    the planet-moon system can stay bound to the neutron star or can leave the remnant and travel as a rogue system.
    The moons stay bound to the planet in both scenarios.
    }
    \label{fig:szepabra}
\end{figure}

We ran the simulation and analysis of 4412 models in four different sets.
In the fiducial set of models, both the planets and the moons were placed on circular orbits.
In the second and third sets of models, the eccentricity of the planet and that of the moon were increased, respectively.
As a fourth set of models, we examined an arrangement where two moons were orbiting a planet in mean-motion resonance.
The parameters used in the SN~II simulations are visualized in Fig.~\ref{fig:szepabra} and summarised in Table~\ref{tab:params}.

\textbf{Stellar parameters:}
Our preliminary results suggested that the mass of the SN~II ejecta has a negligible effect on the eccentricity excitation of the moons.
Here, the stellar envelope mass was chosen to be $M_{\mathrm{ej}}=8~M_\odot$.
In our previous works, we also found that neutron star mass has little to no influence on the perturbation of the companions' orbital elements \citep{regalyetal22, frohlichetal23}.
Therefore, we adopted an intermediate neutron star mass, namely $M_{\mathrm{n}}=2~M_\odot$. 
The resulting total mass of the star pre-explosion was thus $M_{\mathrm{ej}}+M_{\mathrm{n}}=10~M_{\odot}$.
Although this is among the smallest of the SN~II progenitors, it is also one of the most common according to the initial mass function of stars \citep{salpeterxx}.
The radius of the progenitor was set to $500~R_\odot$ in all models based on the lower radius limit given by \citet{Irani2024ApJ...970...96I}.
We adopted an ejecta velocity of $v_{\mathrm{max}}= 1,000~\mathrm {km~s^{-1}}$ and $10,000~\mathrm {km~s^{-1}}$ based on the work of \citet{hamuy-pinto}.

\textbf{Planet parameters:}
We assumed that the companions orbit outside the stellar envelope pre-explosion, as it is still debated whether a body can survive within the envelope of the progenitor \citep{setiawan-etal, bear-etal, chamandyetal, lagosetal, szolgyenetal}.
The planets' semi-major axis was sampled in the range of $a_{\mathrm{p}}^{(0)} \in [5.14: 200]$~au.
This means that the stellar ejecta passes by the planets before they make a full orbit around the progenitor.
In the fiducial set of models, the eccentricity of the planets was chosen to be zero, $e_{\mathrm{p}}^{(0)}=0$.
Models were also run where the planets orbited in eccentric orbits, $e_{\mathrm{p}}^{(0)} \in [0.1:0.8]$.
In almost all cases, the planet is placed at the pericentre of its orbit at the moment of explosion (i.e. its true anomaly is $f_{\mathrm{p}}^{(0)}=0^{\circ}$).
However, in the models where the planet is assumed to orbit the SN progenitor on an eccentric orbit, we have also investigated true anomalies of $f_{\mathrm{p}}^{(0)}=90^{\circ}, 180^{\circ}$ and $270^{\circ}$ as well.
We modelled planets from super-Earths to gas giants, so their masses are $M_{\mathrm{p}} \in [0.01:10] ~M_{\mathrm{Jup}}$.

\textbf{Moon parameters:}
We assumed that the moon-to-planet mass ratio is constant in all cases.
Adopting the mass ratio of Europa relative to Jupiter means that, in units of planetary mass, $M_\mathrm{m}=M_{\mathrm{Eur}}/M_{\mathrm{Jup}}=2.5\times 10^{-5}$.
We also adopted the orbital distance of Jupiter's Galilean moons in units of planetary radius, meaning that $a_{\mathrm{m}}^{(0)}=[5.89; 9.39; 14.97; 26.33]~R_{\mathrm{p}}$.
The pre-explosion true anomaly of the moons, $f_{\mathrm{m}}^{(0)}$, was sampled at $0^\circ,~ 90^\circ,~ 180^\circ,$ and $270^\circ$.
The pre-explosion eccentricity of the moons, $e_{\mathrm{m}}^{(0)}$, is set to zero in the fiducial set of models.
However, models were run in which the orbits of the moons are eccentric, $e_{\mathrm{m}}^{(0)}\in [10^{-4}:10^{-1}]$. 

In the Solar System, there are several pairs of moons or systems of moons that are in mean-motion resonance with each other (e.g., Io, Europa, and Ganymede around Jupiter, or Dione and Enceladus, Tethys and Mimas, Hyperion and Titan around Saturn).
Furthermore, numerical simulations by \citet{ogiharaida12} have shown that the existence of a 2:1 resonance should be common between exomoons.
As such, we also modelled systems in which two moons are in a mean motion resonance based on the example of Io, Europa, and Ganymede, which orbit in a 4:2:1 resonance.
In the resonant systems investigated, the inner moon was orbiting at Europa's distance, with a mean motion of $n_{\mathrm{m}}^{(0)}$ pre-explosion.
The outer moon (with a mean motion of $n_{\mathrm{res}}^{(0)}$) was then placed in a 1:2, 1:3 or 2:3 resonance.
According to the work of \citet{tokadjianpiro23}, eccentricity excitation can depend on the mass ratio of the moons, so we varied the mass of the resonant outer body as $M_{\mathrm{res}}=[0.5;1;2] ~M_{\mathrm{m}}$.

\section{Results}
\label{sec:results}

In general, the SN~II explosion causes the planets' orbits to become unstable, resulting in them continuing their journey through the galaxy as rogue planets.
The only exception to this is models where the planet is located at the apocenter of a highly eccentric orbit at the moment of explosion, see details in Sect.~\ref{sec:results:eccp}.
However, the moons always remain bound to their planets in elliptical orbits.
In the following, we will examine in detail the results of the four sets of models.

\begin{figure}[ht!]
    \centering
    \includegraphics[width=0.9\columnwidth]{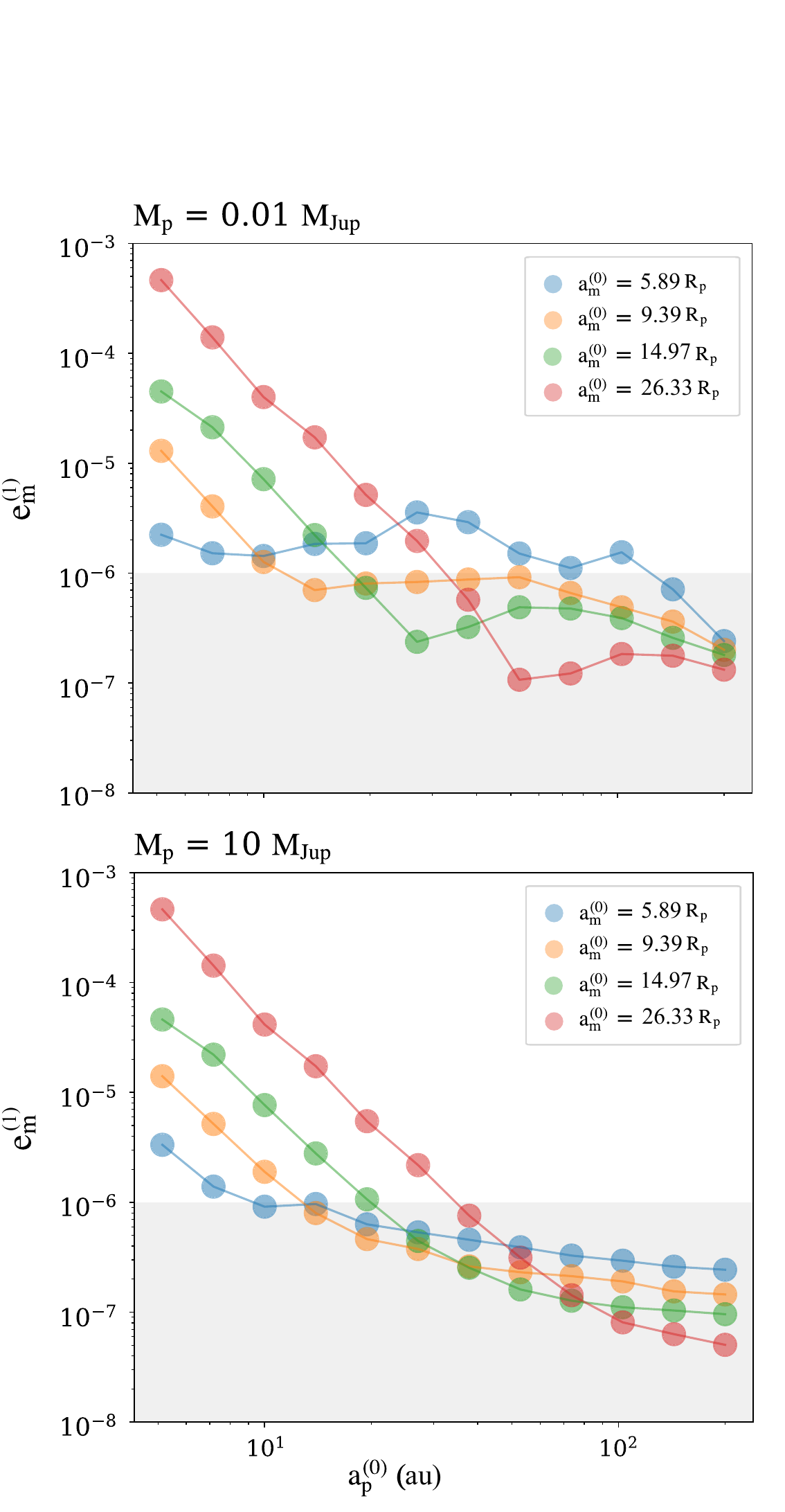}
    \caption{
    The eccentricity of the moons as a function of their planet's pre-explosion semi-major axis, assuming two planetary masses.
    Colours represent pre-explosion planet-moon distances. Only models with $f_{\mathrm{m}}^{(0)}=0^\circ$ are shown as the effect of the pre-explosion true anomaly of the moons is negligible.
    Grey areas mark eccentricity values that arise from perturbations inherent to the hierarchical three-body problem.
    }
    \label{fig:em_ap_nu_amoon}
\end{figure}

\subsection{Fiducial set of models}
\label{sec:results:base}

In the fiducial case, where both planets and moons have circular orbits pre-explosion, we find that the eccentricity of the moons increases in 63\% of the cases.
The maximum eccentricity of the moons in the fiducial set of models is $7.3\times 10^{-4}$.
In the remaining 37\%, the orbital eccentricity is commensurable to the minimum eccentricity inherent in hierarchical triple systems.
The planet-moon distance in this set of models changes by a negligible amount, namely $\lesssim 0.02\%$.

Figure~\ref{fig:em_ap_nu_amoon} shows the post-explosion eccentricity of the moons as a function of the pre-explosion semi-major axis of the planet in two scenarios,  assuming two different planetary masses, namely 0.01 and 10 $M_{\mathrm{Jup}}$.
In each panel, colours correspond to different planet-moon distances, while symbols represent the pre-explosion true anomaly of the moons.
The closer the planet orbited the SN progenitor, the larger the post-explosion eccentricity of the moons.
Accordingly, eccentricities of the order of $10^{-4}$ can only be obtained at the innermost orbits, 5.14~au and 7.17~au.
Assuming a higher mass for the planet has little to no effect on the moons' post-explosion eccentricity.
The effect of the true anomaly of the moons is also negligible.
The reasons for this are discussed in more detail in Sect.~\ref{sec:disc:dynamics}.

Planet-moon distance strongly influences the post-explosion eccentricity:
The eccentricity of the moons decreases as they orbit closer to the planet, as a moon orbiting closer is more deeply embedded in the planet's potential well.
The above statement is not valid for moon eccentricities below $10^{-6}$, which is the inherent eccentricity of the investigated hierarchical three-body systems. 
Repeating the same analysis with different SN ejecta velocities yields plots qualitatively similar to Fig.~\ref{fig:em_ap_nu_amoon}, as the post-explosion moon eccentricity is largely insensitive to $v_{\mathrm{max}}$ within the investigated parameter space.

The velocity of the rogue planet-moon systems is most heavily influenced by the pre-explosion star-planet separation:
closer planets left the system with a higher velocity post-explosion.
The effect of ejecta velocity is negligible, and can only be detected for planets orbiting at $a_{\mathrm{p}}^{(0)}\lesssim15~\mathrm{au}$.
Overall, the rogue planet-moon systems' velocity is between 6 and $32~\mathrm{km~s^{-1}}$.
This means that such rogue planets are still bound to the Galaxy, as the escape velocity of the Milky Way in the Solar neighbourhood is $500-550~\mathrm{km~s^{-1}}$ \citep{mwescape}.

Although the post-explosion velocity of the remnant neutron stars increases with the mass of the planets, their velocities are only between $6\times10^{-6}$ and $4\times10^{-2}~\mathrm{km~s^{-1}}$, which in contrast to our previous work \citep{regalyetal22, frohlichetal23}.
For more details on the peculiar velocity of the rogue planet-moon systems and neutron stars, see Sect.~\ref{sec:disc:dynamics}. 

\subsection{Eccentric planets and moons}
\label{sec:results:eccp}

\begin{figure}[t!]
    \centering
    \includegraphics[width=0.95\linewidth]{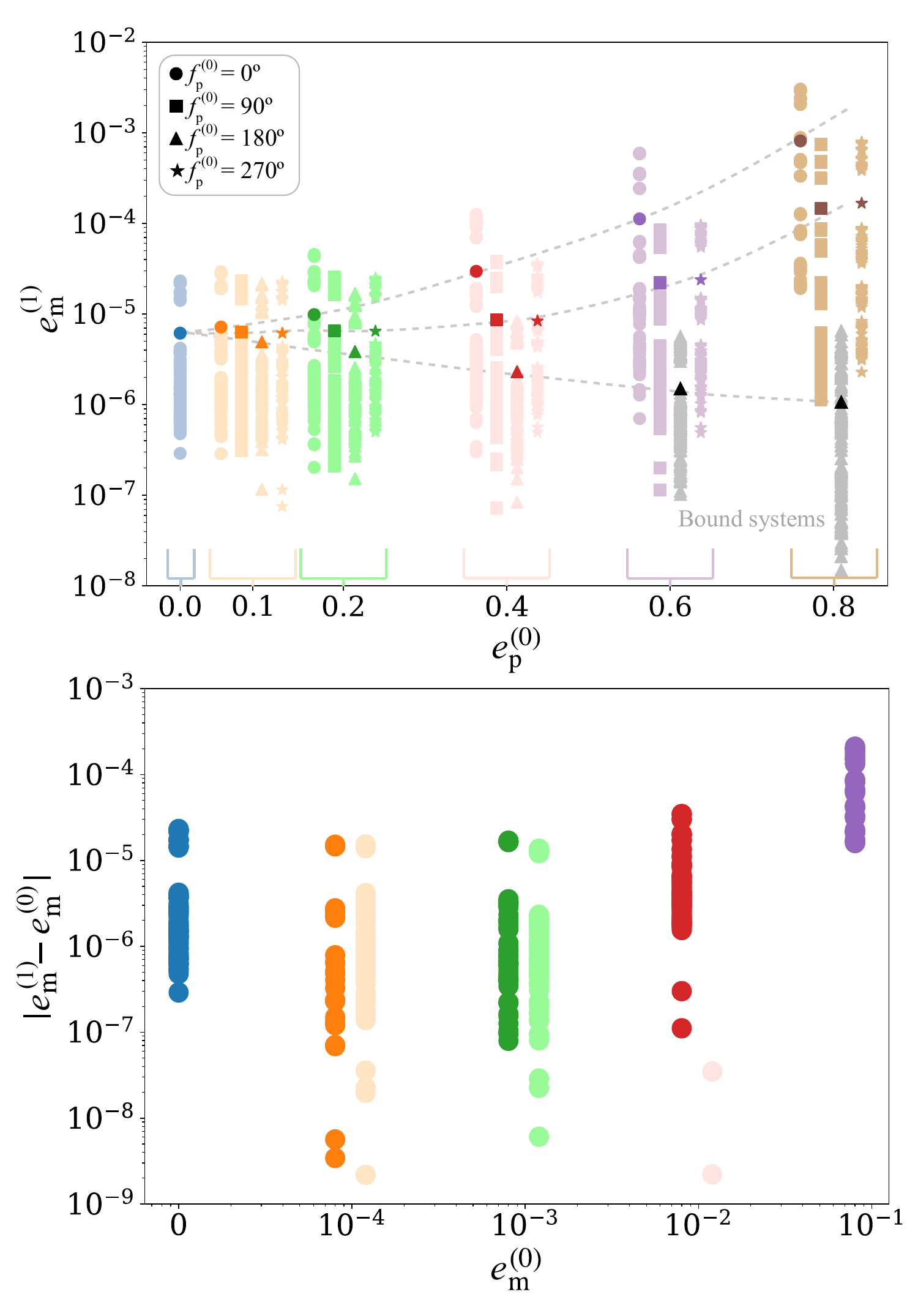}
    \caption{
    Top: eccentricity of moons post-explosion as a function of the pre-explosion eccentricity of the planet.
    Colours represent the pre-explosion eccentricity of the planet, while symbols refer to its pre-explosion true anomaly.
    Systems where the planets stay bound to the remnant neutron star (and hence do not become rogue planets) are indicated in gray.
    The mean values of the moon eccentricities for a given planetary eccentricity-true anomaly combination are indicated by vivid-coloured symbols.
    Dashed lines connect the mean moon eccentricities as a function of planetary eccentricity for the different pre-explosion true anomalies.
    Bottom: change in the eccentricity of moons post-explosion as a function of their pre-explosion eccentricity.
    Models with increasing eccentricity are shown in bright colours, while those with decreasing eccentricity are shown in pastel colours.
    }
    \label{fig:eccp_em_ep0}
\end{figure}

In the second set of models, we assumed the planets to have pre-explosion eccentricities of [0.1:0.8].
Their semi-major axis is chosen to be 13.95~au, as this is the smallest orbit with a pericentre larger than the SN~II progenitor even at $e_p^{(0)}=0.8$ in the fiducial set of models.

The top panel of Fig.~\ref{fig:eccp_em_ep0} shows the post-explosion eccentricity of the moons in relation to the pre-explosion eccentricity and pre-explosion true anomaly of the planets (shown by different colours and symbols, respectively), each point corresponding to a different set of initial conditions.
It can be seen that, for a given pre-explosion planetary eccentricity, the highest post-explosion moon eccentricities are reached when the planet is at the pericentre at the moment of explosion ($f_p^{(0)} = 0^\circ$), and the lowest when it is at the apocentre ($f_p^{(0)} = 180^\circ$).
This is reflected in the mean of the eccentricities, derived from the models with the same $e_\mathrm{p}^{(0)}$  and $f_\mathrm{p}^{(0)}$ and plotted with vivid-coloured symbols.
The values $f_p^{(0)} = 90^\circ$ and $f_p^{(0)} = 270^\circ$ produce similar moon eccentricities, which is due to the fact that the planetary orbit is symmetric to the major axis.
Increasing $e_{\mathrm{p}}^{(0)}$ increases the mean moon eccentricity in the $f_p^{(0)} = 0^\circ$ case.
On the contrary, in the $f_p^{(0)} = 180^\circ$ case, $e_\mathrm{m}^{(1)}$ decreases with $e_{\mathrm{p}}^{(0)}$.
The mean of $e_{\mathrm{m}}^{(1)}$ also increases with $e_{\mathrm{p}}^{(0)}$ in both the $f_p^{(0)} = 90^\circ$ and $f_p^{(0)} = 270^\circ$ cases, however, the effect is not as pronounced as in the $f_p^{(0)} = 0^\circ$ case.

The highest post-explosion moon eccentricity is reached when $e_p^{(0)} = 0.8$ and $f_p^{(0)} = 0^\circ$, and is found to be $3\times10^{-3}$. 
This is an order of magnitude larger than what was found for planets in circular orbits.
The semi-major axis of the moons in this set of models changes by $\lesssim 0.2\%$.

It is important to note that in the $e_{\mathrm{p}}^{(0)}=0.6$ and $e_{\mathrm{p}}^{(0)}=0.8$ cases, the planet can remain bound to the remnant neutron star (indicated by gray symbols in the top panel of Fig.~\ref{fig:eccp_em_ep0}).
Nevertheless, the moons also remain bound to these planets, suggesting that exomoons may also orbit pulsar planets.
Characterizing such systems requires further simulations on an extended parameter space.

As a third set of models, we investigated the eccentricity evolution of moons with non-circular pre-explosion orbits.
The planet is placed at $a_p^{(0)}=13.95~\mathrm{au}$ on a circular orbit, while the eccentricity of the moons is assumed to be $10^{-4}$, $10^{-3}$, $10^{-2}$, or $10^{-1}$.

The bottom panel of Fig.~\ref{fig:eccp_em_ep0} shows the absolute value of the change in eccentricity of the moons, grouped by their pre-explosion eccentricity.
The change in eccentricity can be positive or negative, shown by bright and pastel colours, respectively.
For the models with $e_{\mathrm{m}}^{(0)}\gtrsim10^{-2}$, the post-explosion eccentricity increases.
The same is naturally true for $e_{\mathrm{m}}^{(0)}=0$.
The maximum change in the post-explosion moon eccentricity is $3\times10^{-4}$.
This means that if a moon was already in an eccentric orbit pre-explosion, it will keep its eccentricity post-explosion.
The planet-moon distance in this model set changes by a negligible $\lesssim 0.001\%$.

\subsection{Resonant moon systems}
\label{sec:results:resonant}

In the following, we present results regarding the effects of the  SN~II explosions on systems of two moons in 1:2, 1:3, and 2:3 mean motion resonance.
The inner moon (like Jupiter's moon Europa) orbits at a distance of 9.39~$R_{\mathrm{p}}$ from the planet, while the outer moon orbits according to the resonance.

The resonant configurations were constructed by placing the inner moon at a fixed semi-major axis and determining that of the outer moon such that the ratio of their mean motions exactly satisfied the integer commensurability 
$n_1:n_2=p:q$. This ensures the system lies at the nominal location of the mean-motion resonance. We did not apply convergent migration or eccentricity damping forces (e.g. \citealp{tamayo-etal-17,lammers-winn-24}) to achieve libration of the resonant angles, since our aim was to study the dynamical consequences of supernova mass loss for already resonant -- or near -- resonant-systems rather than to model resonance capture or long-term resonant stability.

We find that the resonant configuration changes only negligibly due to mass loss in the SN~II explosion.
This is because, similarly to the single-moon systems, the semi-major axis of the moons remains practically unchanged (the change is $\lesssim 0.2\%$).
The change of resonance is most pronounced in the 2:3 and 1:3 resonances ($\lesssim1\%$), while it is the weakest in the 1:3 resonance ($\lesssim0.1\%$).
Overall, we conclude that the moons remained in resonance post-explosion.

As the relative position of the two moons changes, non-radial forces are introduced, and a circular orbit is no longer possible. 
Consequently, the eccentricity of the moons is excited.
Our simulations show that the maximum eccentricities induced by resonance are approximately $3\times10^{-4}$ for the 1:3 resonance, and $2\times10^{-2}$ for both the 1:2 and 2:3 resonances.
In all cases, the outer moon acquires an eccentricity comparable in magnitude to that of the inner moon due to the resonant coupling.
The additional eccentricity induced by the SN~II explosion is negligible, as it is orders of magnitude smaller than the intrinsic excitation caused by the resonant configuration (see Fig.\ref{fig:eccp_em_ep0}).
The eccentricities arising from resonance are nearly an order of magnitude greater than those found in rogue planet-moon systems with eccentric pre-explosion planetary orbits, and several orders of magnitude larger than those in our fiducial (circular planetary orbit, singular moon) model set.

\section{Discussion}
\label{sec:disc}

\subsection{Dynamical considerations}
\label{sec:disc:dynamics}

Based on our results, planets ejected during SN~II explosions retain their moons in all cases.
This aligns with several studies showing that rogue planets ejected by early dynamical instabilities also retain their moons \citep{debessigurdsson07, hongetal18, rabagosteffen18}.

As the expanding SN envelope reaches the planet–moon system, the eccentricity of the moon increases. 
This arises from the substantial mass loss during the explosion, which alters the gravitational potential and induces changes in the acceleration - and thus the velocity - of both the remnant neutron star and the planet.
The ejected envelope retains the progenitor star’s velocity at the moment of explosion. 
Meanwhile, the neutron star and the planet receive velocity kicks due to momentum conservation, resulting in non-zero peculiar velocities even though the supernova is assumed to be spherically symmetric (see further detail in \citealp{regalyetal22}). 
The magnitude of the moon’s eccentricity excitation depends solely on the velocity kick received by the planet.

\begin{figure}[t!]
    \centering
    \includegraphics[width=\columnwidth]{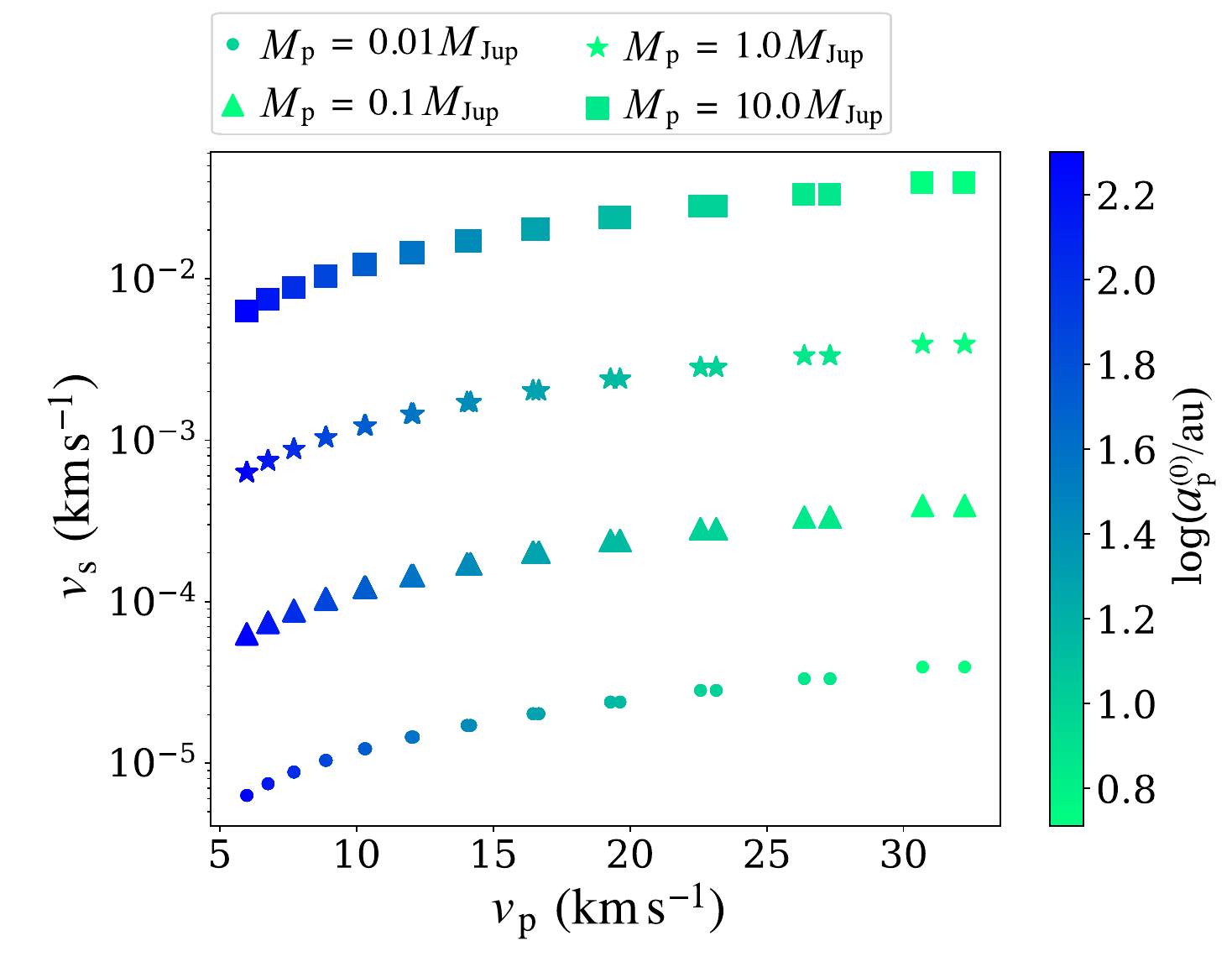}
    \caption{Post-explosion peculiar velocities of planets and remnant neutron stars in the fiducial set of models.
    Colours represent the pre-explosion semi-major axis of the planets, while symbols correspond to different planetary masses.}
    \label{fig:vpec}
\end{figure}

We found that the closer a planet orbits its progenitor star, the greater the increase in the eccentricity of its moons post-explosion (see Sect.\ref{sec:results:base}). 
Figure~\ref{fig:vpec} shows the post-explosion peculiar velocities of the remnant neutron stars as a function of those of the planet–moon systems in our fiducial model set, with different planetary semi-major axes indicated by colours and planetary masses by symbols. 
The figure clearly illustrates that planets initially located closer to the progenitor acquire higher peculiar velocities. 
These peculiar velocities directly reflect the magnitude of the velocity kick imparted during the SN~II explosion. 
Given that the excitation of moon eccentricity depends solely on the change of planetary velocity, moons orbiting close-in planets acquire higher eccentricities after the explosion.

Fig.~\ref{fig:vpec} clearly shows that the post-explosion peculiar velocity of the remnant neutron stars is most heavily influenced by the mass of the planet:
more massive planets result in higher remnant velocities. 
In our fiducial models, the peculiar velocities of the remnants lie between $6\times10^{-6}~\mathrm{km~s^{-1}}$ and $4\times10^{-2}~\mathrm{km~s^{-1}}$.
This is in stark contrast to our earlier findings, where stellar velocities reached a few hundred $\mathrm{km~s^{-1}}$ \citep{regalyetal22, frohlichetal23}.
The difference appears because such high velocities require the SN progenitor to have been part of a binary system with a stellar-mass companion. 

The growth in a moon’s eccentricity depends only weakly on its pre-explosion true anomaly (see Sect.~\ref{sec:results:base}).
The smallest change occurs at $f_{\mathrm{m}}^{(0)} = 0^\circ$, when the moon's orbital velocity is aligned with the supernova-induced kick velocity of the planet.
The largest change occurs at apocentre ($f_{\mathrm{m}}^{(0)} = 180^\circ$), while the intermediate pre-explosion positions ($f_{\mathrm{m}}^{(0)} = 90^\circ$ and $270^\circ$) produce intermediate eccentricity excitations.
Overall, this effect is negligible in our simulations because the moon’s orbital speed is much smaller than the kick received by the planet.

In this study, we assumed that the planet-moon orbit is coplanar with the planet-star orbit (i.e, $i_{\mathrm{m}}^{(0)} = 0^\circ$).
Additional simulations with pre-explosion moon inclinations of $i_{\mathrm{m}}^{(0)}=5^\circ$, $7^\circ$, $10^\circ$, $45^\circ$, and $90^\circ$ indicate that the inclination of the moons is unaffected by the SN~II explosion (the maximum change in inclination is found to be around 0.01\%).
Furthermore, by comparing these simulations with cases where $i_m^{(0)} = 0$, we find that increasing the inclination of the moons does not affect the qualitative outcome of the simulations with regard to the post-explosion eccentricity or semi-major axis of the moons.
As we have seen, the effect of the true anomaly is minimal at $i_m^{(0)} = 0$. 
As the inclination increases, the effect of the true anomaly begins to disappear and is completely eliminated at $90^\circ$ since the velocity vector of the moon is always perpendicular to the kick received by the planet, regardless of its true anomaly. 
This explains why the effect of inclination is also negligible.

For circular planetary orbits, the closer a planet orbits its SN~II progenitor, the higher the post-explosion eccentricity of its moon, as shown in Fig.~\ref{fig:em_ap_nu_amoon}. 
When the planet’s pre-explosion orbit is eccentric, the resulting post-explosion moon eccentricity also depends strongly on the planet’s true anomaly at the moment of explosion (see Fig.~\ref{fig:eccp_em_ep0}). 
If the planet is at pericentre, the moon’s post-explosion eccentricity increases with the planet’s orbital eccentricity. 
Conversely, if the planet is at apocentre, the moon’s eccentricity decreases as planetary eccentricity increases. 
This can be explained by the following chain of thought.
At a constant semi-major axis for different eccentricities, orbital velocity increases with eccentricity at pericentre and decreases at apocentre.
The peculiar velocity of rogue planets is proportional to their pre-explosion orbital velocity (see Fig.~\ref{fig:vpec}), which, in turn, sets the strength of the moon's eccentricity excitation (see Fig.~\ref{fig:em_ap_nu_amoon}). 
Thus, post-explosion moon eccentricities will be highest if the planet was at the pericentre of an eccentric orbit at the moment of explosion, and lowest if it was at the apocentre.
Note, however, that planets on highly eccentric orbits ($e_{\mathrm{p}}^{(0)} = 0.6$ or $0.8$) may not escape from the system, and thus remain gravitationally bound to the remnant neutron star, which is consistent with our earlier work of \citet{regalyetal22}.

The semi-major axes of the moons experience only negligible changes -- at most 0.2\% -- as a result of the SN~II explosion. 
Given this near-constancy, any existing resonant configurations are preserved post-explosion, as demonstrated in Sect.~\ref{sec:results:resonant} and consistent with the findings of \citet{rabagosteffen18}. 
Therefore, for moon pairs initially in non-resonant orbits, the minimal change in semi-major axis is insufficient to bring them into resonance.
However, long-term tidal evolution can cause the inner moon to be captured into resonance with the outer moon.

\subsection{Urability on the moons of rogue planets?}

The word habitability is often used to describe celestial bodies where conditions allow for the existence of liquid surface water.
Note that what exactly it means to be habitable and how we should define it is still a matter of debate \citep{lammeretal09}.
However, even if a celestial body is habitable, it does not necessarily bear the conditions for sprouting life.

Urability -- a recently coined term by \citet{deameretal22}, in which the Latin prefix ‘ur-’ means ancient, primitive, or earliest -- is concerned with the question of what are the absolute minimum physical, chemical, and energetic conditions necessary for life to emerge.
As opposed to habitability, it focuses on whether life could have begun on a particular celestial body, and does not concern itself with the long-term evolution and sustainment of that life.
An apt example:
Earth today can naturally be considered habitable.
However, life as we know it could not start in the current $\mathrm{O}_2$-rich, oxidative atmosphere and many other conditions.
On the other hand, 3.5 billion years ago, our Earth was urable, as that was exactly the time when life came to bloom.
However, Earth is considered an uninhabitable planet at this point because of its high temperatures and $\mathrm{CO}_2$-rich, reductive atmosphere.

When looking for urable environments, exomoons seem like a good bet:
based on our own Solar System, exomoons must be plentiful, and the sources of energy are also numerous.
In fact, there are four other sources of heat in addition to direct irradiation:
1) starlight is reflected by the moon's planet; 
2) the planet radiates its internal heat;
3) the moon also harbours geothermal heat, and 
4) if the moon's orbit is eccentric, tidal heating also plays an important role \citep{jacksonetal08}.
Note that most of the above sources are independent of distance from the star, and if the moon has an atmosphere, the energy accumulated may even cause a greenhouse effect \citep{hellerbarnes15}.
Even tidal heating alone can result in a suitable surface temperature for liquid water at large stellar distances \citep{reynoldsetal87, petersturner13, hellerarmstrong14, dobosturner15}.

The contribution of tidal heating to increasing the temperatures of moons is now unquestionable.
Our Solar System provides excellent examples of this in the form of Europa \citep{carretal98}, Enceladus \citep{thomasetal16}, Miranda \citep{miranda}, and Mimas \citep{laineyetal24}. 
Due to their non-negligible eccentricity ($\simeq10^{-2}$), tidal heating is so significant that the water content of the moons can persist in a global liquid subsurface ocean up to tens of kilometres thick under the ice crust \citep{yoder79, porcoetal06}.

Deep-sea hydrothermal vents in the deepest parts of the oceans have played a major role in the development of life on Earth, providing a constant source of energy and nutrients for the extremophilic bacteria that form the basis of the ecosystem \citep{barosshoffman85}.
If life on Earth sprouted here, then such hydrothermal vents can be considered an urable environment.
In light of this, it is not too far-fetched to suggest that tidally heated subsurface oceans of exomoons may provide the basic conditions necessary for life to emerge \citep{dobosturner15, tjoaetal20}.
Furthermore, in the absence of an atmosphere, a thick ice shell above a subsurface ocean also provides effective shielding from cosmic rays and asteroid impacts.
Life in such extreme and isolated environments may be driven to evolve novel adaptations in response to environmental pressures \citep{karvewagner22}.
Nonetheless, random mutations continue to play a central role in evolution, since even small genetic changes can significantly affect which traits are passed on and how they respond to natural selection \citep{giffordetal24}.
Note that a contact between the ocean and a silicate core is also necessary to supply a reservoir of organic compounds \citep{lammeretal09}.

In this study, we assume that the moons lack an atmosphere, following the work of \citet{lammeretal09, lichtenbergeretal10}. 
However, an optically thick atmosphere — along with eccentricities between $10^{-3}$ and 0.5, and orbital distances of $10^{-3}$-$10^{-2}$~au from a Jupiter-mass host planet — could make surface life possible on an Earth-mass exomoon \citep{avilaetal21}. 
Under such conditions, surface oceans might persist for 50 million to 1.6 billion years \citep{roccettietal23}.
Here, however, we focus solely on the potential for life in subsurface environments.

\subsubsection{Tidal heating}
\label{sec:results:tidal}

\begin{figure*}[ht!]
    \centering
    \includegraphics[width=0.9\linewidth]{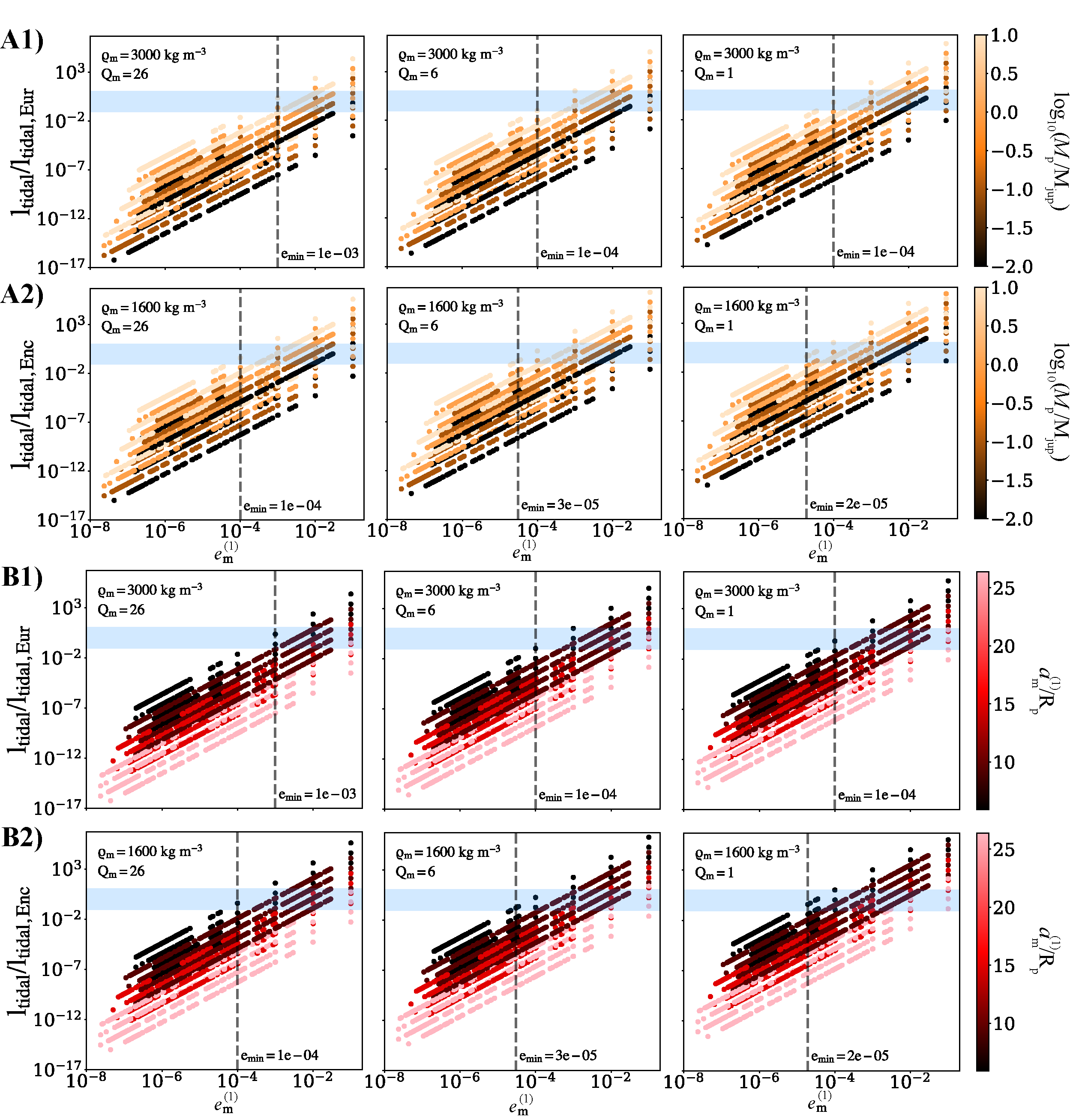}
    \caption{
    Specific power of tidal heating on the moons of rogue planets as a function of the eccentricity of the moons for different values of $\rho_{\mathrm{m}}$ and $Q_{\mathrm{m}}$.
    The colours in panels A1 and A2 represent the mass of the planets, and in panels B1 and B2, the distance of the moons from the planet.
    In the light blue bands, $l_{\mathrm{tidal}}/l_{\mathrm{tidal,Eur}}$ and $l_{\mathrm{tidal}}/l_{\mathrm{tidal,Enc}}$ fall between 0.1 and 10, so these moons can be deemed urable.
    The minimum eccentricities required for urability are indicated by dashed lines.
    }
    \label{fig:ltidal}
\end{figure*}

As we have seen in Sect.~\ref{sec:results}, SN~II explosions can result in moons orbiting rogue planets in stable but eccentric orbits.
Moreover, if they were already eccentric pre-explosion, this will not change after the supernova event.
The latter is true for both single-moon and resonant moon systems.
Even though such moons are no longer heated by the radiation from their host stars, they can be heated substantially by the dissipation of tidal forces.

Here, we apply to the rogue planet-moon systems a constant phase-lag (CPL) model based on the works of \citet{macdonald64, efroimskymakarov13}, discussed further in Appendix~\ref{sec:apx:tidal}.
In this model, tidal luminosity is calculated as
\begin{equation}
    L_{\mathrm{tidal}}= \frac{21}{2} G \frac{k_{2,\mathrm{m}}}{Q_{\mathrm{m}}} M_{\mathrm{p}}^{5/2} R_{\mathrm{m}}^{5} \frac{(e_{\mathrm{m}}^{(1)})^2}{(a_{\mathrm{m}}^{(1)})^{15/2}}.
    \label{eq:cpl}
\end{equation}
We used values of $k_{\mathrm{2,m}}=0.26$, while $Q_{\mathrm{m}}$ is sampled as 1, 6, and 26 – the latter two corresponding to Europa’s minimum and maximum values estimated by \citet{ferrazmello12}.
The specific power of tidal heating is
\begin{equation}
    l_{\mathrm{tidal}} = \frac{ L_{\mathrm{tidal}} }{ A_{\mathrm{m}}},
    \label{eq:kisl}
\end{equation}
where $A_{\mathrm{m}}$ is the surface area of the moon.

The radius of the moons is calculated based on their density, differentiating between two cases:
in one instance the density of the moons is $\rho_{\mathrm{m}}=3000~\mathrm{kg~m^{-3}}$, similar to that of Europa \citep{eurdensity}.
In the other case, we investigated at half of the above density value, $1600~\mathrm{kg~m^{-3}}$, analogous to the mean density of Enceladus \citep{encdensity}. 

Figure~\ref{fig:ltidal} shows the specific power of tidal heating normalised to the values estimated for Europa and Enceladus as a function of the eccentricity of the moons of the rogue planets modelled in Sect.~\ref{sec:results}.
Colours in rows A1 and A2 correspond to different planetary masses, while colours in rows B1 and B2 mark different planet-moon distances.
Rows A1 and B1 assume that the bulk density of the moons is analogous to the bulk density of Europa, while rows A2 and B2 assume bulk densities similar to that of Enceladus.
Here, we consider moons urable if the specific tidal heating power is at least one-tenth and at most ten times the value estimated on Europa or Enceladus\footnote{A strict quantitative criterion for whether a body is urable is difficult to establish. However, a minimal requirement is that the base of the ice shell must melt to allow contact with the silicate core.}.
These limits are marked by light blue bands in each panel.

It can be seen that for each $Q_{\mathrm{m}}$-$\rho_{\mathrm{m}}$ pair (shown in the different panels), there is a minimum eccentricity, $e_{\mathrm{min}}$, at which the specific tidal heating power is one tenth of that measured on Europa or Enceladus.
This $e_{\mathrm{min}}$ grows with the density of the moon and with the value of $Q_{\mathrm{m}}$.
Overall, it is found to be between $2\times10^{-5}$ and $10^{-3}$ in the parameter space studied.
With an eccentricity lower than $e_{\mathrm{min}}$, it is difficult to imagine that tidal heating could melt even a thin layer of ice on the moons.

It should be noted that the eccentricity cannot be arbitrarily high either, because then the tidal heating power would increase to the point where the entire water supply of the moon (including the surface ice crust) could melt and then evaporate very quickly (assuming no surface pressure).
Overall, the fraction of models that can be described as urable varies between 12-15\% depending on the choice of $Q_{\mathrm{m}}$ and $\rho_{\mathrm{m}}$.

The specific power of tidal heating increases with the mass of the planet (see the orange colours of panels A1 and A2).
However, if we look at the bands indicating urability, we see that urability is possible for all values of $M_{\mathrm{p}}$ considered here.
Looking at the red colours in panels B1 and B2 (depicting planet-moon separation), we see that $l_{\mathrm{tidal}}$ is larger for the moons orbiting closer to their planets.
As we can see, moons orbiting the planet at or below 15 planetary radii (shown in the darkest colours) can develop an urable environment.
Models with an eccentricity of 0.1 are especially worth highlighting:
these moons can be urable even if their orbital distance is between 15 and 25 planetary radii.

\subsubsection{Temperature limits for water retention}
\label{sec:disc:urability}

For a moon to be urable post-explosion, it is also necessary that its water supply remains bound to the surface and does not boil away under the irradiation of the progenitor star.
Here we examine urability not only post-explosion, but also in the SN~II progenitor system.

According to \citet{hellerbarnes13}, the total power arriving at the moon's surface, taking into account the direct irradiance of the star ($L_{\mathrm{irr}}$), the radiation reflected by the planet ($L_{\mathrm{ref}}$), the thermal irradiance of the planet ($L_{\mathrm{th}}$), the contribution from the decay of radioactive elements in the moons' interior ($L_{\mathrm{rad}}$), and the tidal heating power ($L_{\mathrm{tidal}}$) reads
\begin{equation}
    L_{\mathrm{tot}} = L_{\mathrm{irr}} + L_{\mathrm{ref}} + L_{\mathrm{th}} + L_{\mathrm{rad}} + L_{\mathrm{tidal}}.
\end{equation}
The planet's thermal radiation power is an order of magnitude lower than the tidal heating power, and calculating the heat from the decay of radioactive material would require a good estimate of what fraction of the moon's material is made up of radioactive elements.
The power of radiation reflected by the planet is only commensurable with tidal heating if the moon orbits a giant planet.
For the above reasons and simplicity, we have only considered the direct irradiation of the star and the tidal heating.
Therefore, we assume that the power that heats the moon pre-explosion is 
\begin{equation}
    L_{\mathrm{m}}^{(0)} = L_{\mathrm{irr}} + L_{\mathrm{tidal}},
\end{equation}
while the post-explosion luminosity, in the absence of stellar irradiation, is 
\begin{equation}
   L_{\mathrm{m}}^{(1)} = L_{\mathrm{tidal}}.
\end{equation}
The luminosity from irradiance is calculated as
\begin{equation}
    L_{\mathrm{irr}} = \frac{\sigma 4 \pi R_*^2 T_*^4}{4 \pi R_{\mathrm{m}}^{2}}~\frac{(1-A)}{16 \pi (a_{\mathrm{p}}^{(0)})^2},
\end{equation}
where $R_*$ and $T_*$ are the radius and the temperature of the star, $A$ is the albedo of the moon (assumed to be 0.8 in all cases following the work of \citealp{encalbedo}), and $\sigma$ is the Boltzmann constant.

For simplicity, stellar parameters are determined as the average over the stellar lifetimes from the PARSEC v2.0 stellar evolution model dataset  \citep{costaetal19a, costaetal19b, nguyenetal22}.
Taking the average of these values is supported by the fact that $10~M_\odot$ stars spend an order of magnitude less time on the giant branch than their main-sequence lifetimes ($\lesssim10^6$~yrs versus $\simeq10^7$~yrs).
We adopted the stellar parameters of the 10 solar mass models (same as for the SN~II modelling).
It is worth noting, however, that between 8 and 11.5 solar masses, the luminosities of stars differ only slightly, so that all results obtained here can be extended to these masses.

Following \citet{tokadjianpiro23}, the effective temperature of the moons is obtained by applying the Stefan-Boltzmann law, meaning that pre-explosion
\begin{equation}
    L_{\mathrm{m}}^{(0)} = L_{\mathrm{irr}}+L_{\mathrm{tidal}}= \sigma A_{\mathrm{m}} (T_{\mathrm{m}}^{(0)})^4,
\end{equation}
while post-explosion 
\begin{equation}
    L_{\mathrm{m}}^{(1)} =L_{\mathrm{tidal}}= \sigma A_{\mathrm{m}} (T_{\mathrm{m}}^{(1)})^4.
\end{equation}
As water reservoirs should not be evaporated pre-explosion, we require that $T_{\mathrm{m}}^{(0)}<373~\mathrm{K}$.
Furthermore, urability requires the presence of a liquid subsurface ocean post-explosion, meaning that $T_{\mathrm{m}}^{(1)}>273~\mathrm{K}$.

We assumed that the moons have the same tidal heating parameters as Europa, meaning that $A=0.8$, $Q_{\mathrm{m}}=16$, $k_{\mathrm{2,m}}=0.26$ \citep{eurdensity, ferrazmello12}.
The semi-major axis of the planet is sampled between 10 and 1000~au, and its mass between 0.01 and 10 jupiter masses.
The distance of the moon from the planet is between 4 and 50 planetary radii, and the eccentricity is sampled between $10^{-6}$ and $0.9$.

\begin{figure}[ht!]
    \centering
    \includegraphics[width=0.99\linewidth]{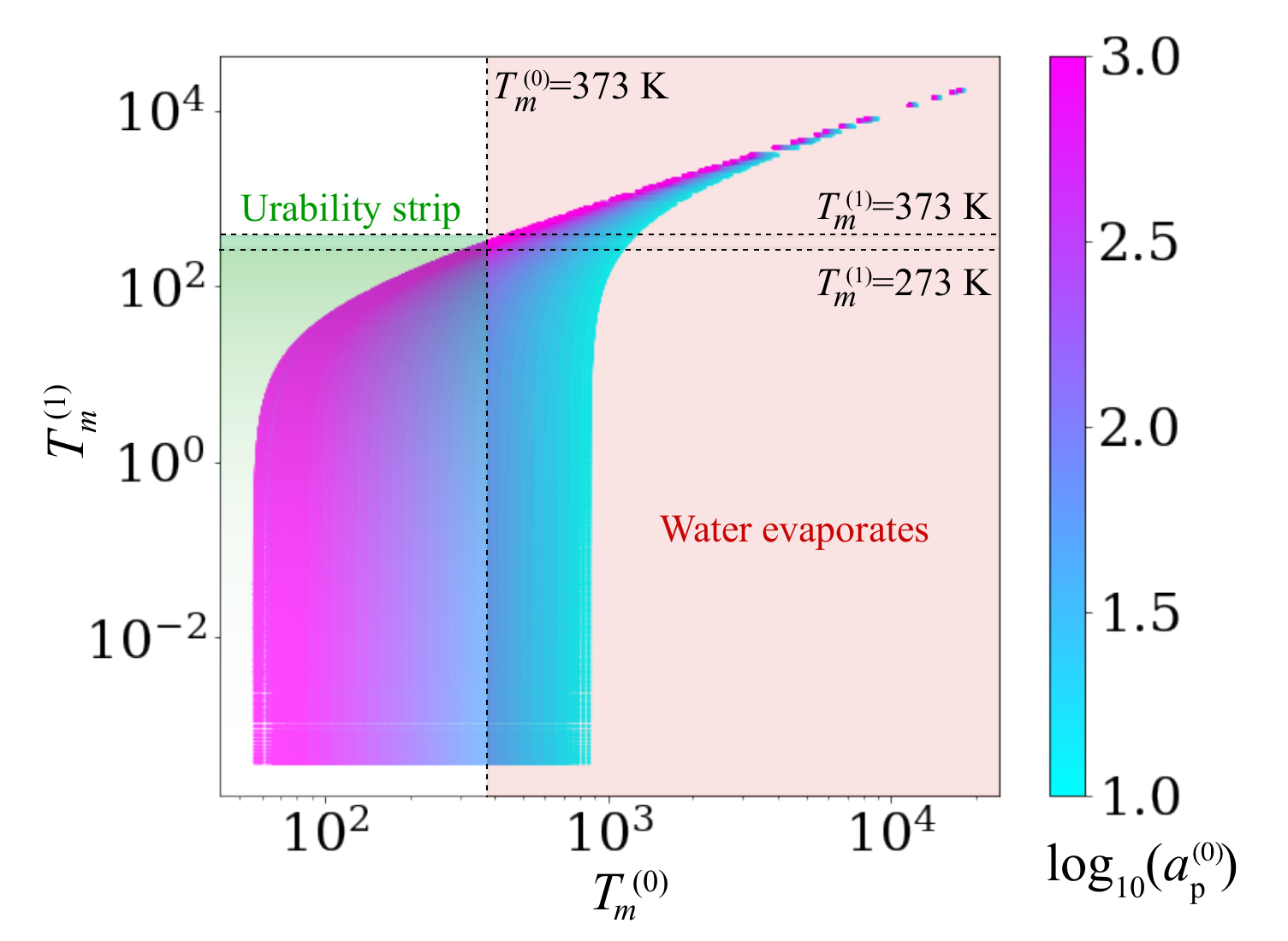}
    \caption{
    The effective temperature of the moons post-explosion as a function of their effective temperature pre-explosion
    The different colours indicate the pre-explosion distance of the planet from the star.
    For the moons in the region marked with red, urability is not a possibility as their water reservoirs have evaporated pre-explosion.
    Moons belonging to the green shaded region, however, might be urable post-explosion, depending on their composition.
    }
    \label{fig:t_tot_ap0}
\end{figure}

Figure~\ref{fig:t_tot_ap0} shows the post-explosion effective temperature of the moons as a function of their pre-explosion effective temperature ($T_{\mathrm{m}}^{(0)}$ and $T_{\mathrm{m}}^{(1)}$, respectively) colour color-coded by the pre-explosion semi-major axes of the planets.
Moons that fall on the right side of the plot (marked in light red), where $T_{\mathrm{m}}^{(0)}>373~\mathrm{K}$, can not be urable as their water content is evaporated pre-explosion.
However, on the left side of the plot, there exists a strip of potential urability (marked by a light green band), where the effective temperature of the moons is between 273~K and 373~K post-explosion, while they are never heated so much by stellar irradiation as to lose their water content pre-explosion.

We note that in the absence of an atmosphere, water can boil at temperatures well below 373~K. 
This evaporation may lead to the transient formation of a thin atmosphere. 
However, it remains an open question how massive a moon must be for its atmosphere to persist long enough to allow surface life to emerge.
Accurately modelling such processes would, however, require a dedicated thermal-atmosphere model, which lies beyond the scope of this study.

The planets of potentially urable moons have a pre-explosion semi-major axis $\geq383$~au, and a mass between 0.1 and 10~jupiter masses.
According to the \href{https://exoplanetarchive.ipac.caltech.edu}{NASA Exoplanet Archive}, the planets discovered so far orbiting at semi-major axes of several hundred au are all at least two Jupiter masses.
The furthest orbiting planets so far have semi-major axes of more than a thousand au.
This argument also favours that urable habitats can emerge beyond the snow line of solar systems \citep{williamsetal97, tjoaetal20}.
Based on our results, the urable moons should orbit between 4 and 36 planetary radii, with an eccentricity greater than 0.1.

In our model, $T_{\mathrm{m}}^{(1)}$ gives an effective temperature, assuming that the moon is homogeneous and tidal heating is effective over its entire volume.
In reality, however, urability can be achieved by the melting of even a thin layer of ice, a few tens of kilometres thick.
Such an inner layer of ice may even have a temperature higher than the values given here (see the case of Europa, \citealp{eurinnertemp}).
If we want to calculate the temperature of the layers of ice and water within the moons, we need to know the stratification of their differentiated structure, which requires a more complex rheological and heating model beyond the scope of this paper.
We can argue, however, that if only a thin layer of ice is assumed to melt due to tidal heating, the data points shown in Fig.~\ref{fig:t_tot_ap0} would be shifted to the upper right.

\subsection{Eccentricity damping of the moons}

The eccentricity of the moons of rogue planets will decrease over time due to tidal dissipation. 
This effect can only be countered if the moon orbits in a resonant configuration with another moon, whereby eccentricity is maintained during the existence of the resonant configuration \citep{tokadjianpiro23}.

Here we discuss the temporal evolution of eccentricity based on the work of \citet{yoderpeale81} and the constant phase-lag model.
In a single-moon, non-resonant system, the damping time scale is
\begin{equation}
    \tau_{\mathrm{e}} = \frac{4}{63} M_{\mathrm{m}} \left (\frac{a_{\mathrm{m}}^{(1)}}{R_{\mathrm{m}}}\right )^5 \frac{\tilde{\mu}_{\mathrm{m}} Q_{\mathrm{m}}}{n_{\mathrm{m}}^{(1)}},
    \label{eq:taue}
\end{equation}
where $\tilde{\mu}_{\mathrm{m}}$ is a measure of the ratio of elastic to gravitational forces at the boundary of the solid core of the moon, assumed to be equal to that of Europa, i.e. $\tilde{\mu}_{\mathrm{m}}=80$ following \citet{mutilde}.
Assuming a constant moon-to-planet mass ratio, $M_{\mathrm{m}}=2.5\times10^{-5}$, after some trivial algebra, we get
\begin{equation}
    \tau_{\mathrm{e}} = 
    \frac{4}{1323} 
    \frac{1}{\sqrt{\pi G}}
    \tilde{\mu}_{\mathrm{m}} Q_{\mathrm{m}} 
    \frac{M_{\mathrm{m}}^{-2/3}}{(1+M_{\mathrm{m}})^{1/2}} 
    \hat{a}^{13/2} 
    \frac{\rho_{\mathrm{m}}^{5/3}}{\rho_{\mathrm{p}}^{13/6}},
    \label{eq:taue_mod}
\end{equation}
where $\hat{a}$ is the post-explosion planet–moon separation measured in the radius of the planet, and $\rho_{\mathrm{p}}$ and $\rho_{\mathrm{m}}$ are the density of the planet and the moon, respectively.
We assumed that the planet and moon densities are equal to those of Jupiter and Europa (i.e. 1326~$\mathrm{kg~m^{-3}}$ and 3031~$\mathrm{kg~m^{-3}}$, respectively).

\begin{figure}[t!]
    \centering
    \includegraphics[width=0.99\linewidth]{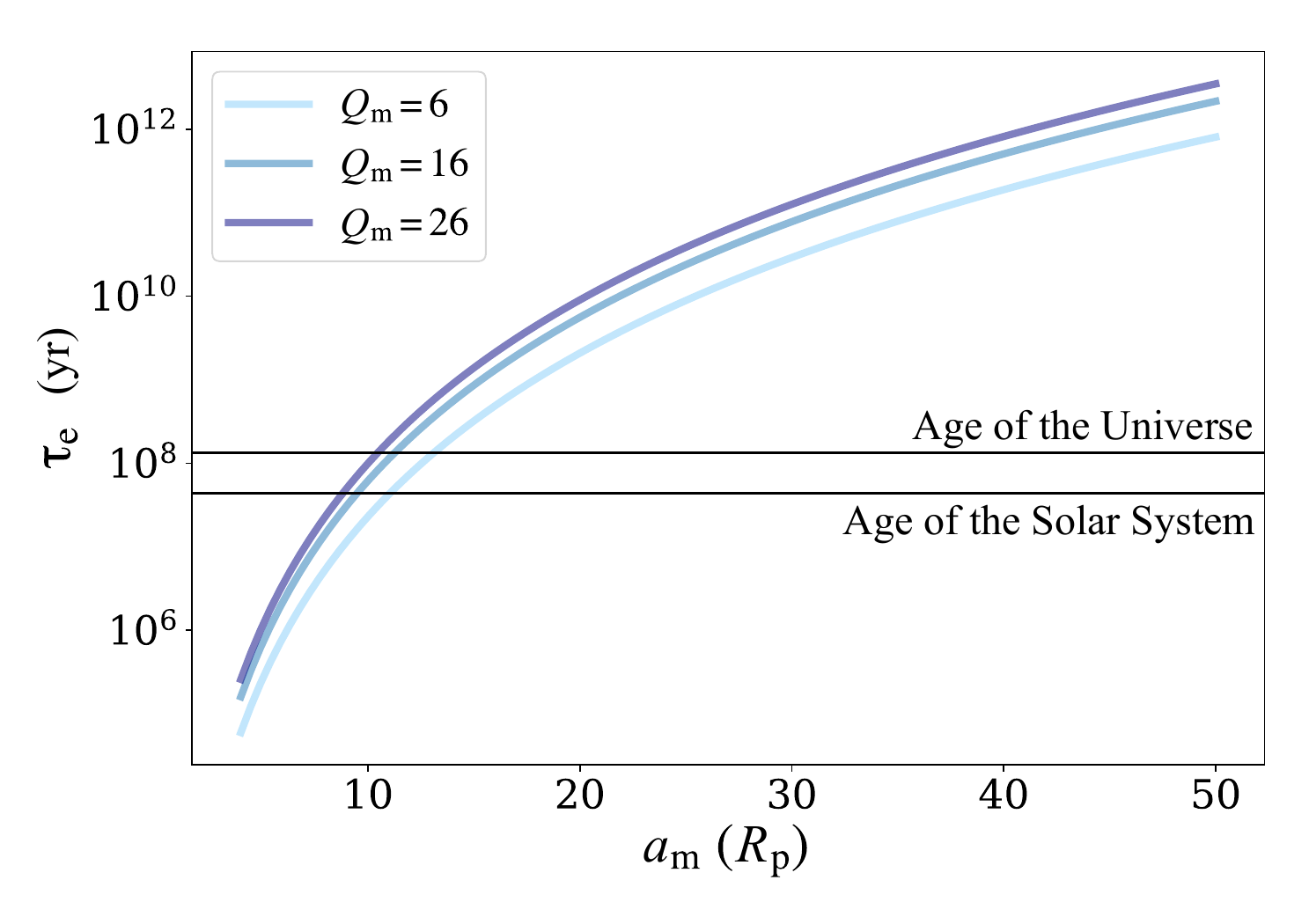}
    \caption{
    Eccentricity dampening timescale of the moons of rogue planets at different planet-moon separations and tidal dissipation functions ($Q_{\mathrm{m}}=6, 16$, and 26), calculated in the CPL model.
    The black lines mark the ages of the Solar System and the Universe.
    }
    \label{fig:eccdamping}
\end{figure}

Figure~\ref{fig:eccdamping} shows the damping timescale for different planet-moon separations and tidal dissipation functions.
The moons orbit at 5-50 planetary radii, an interval that extends the parameter space used in the SN~II simulations.
The resulting damping timescales range from $5\times10^4$ years to $10^{12}$ years.
For moons with $a_{\mathrm{m}}\gtrsim 10~R_{\mathrm{p}}$, $\tau_{\mathrm{e}}$ exceeds the age of the Solar System, and if the moons orbit beyond $\simeq 13$ planetary radii, the damping timescales are comparable to the age of the Universe.
In contrast, it took only about 0.7 billion years for life on Earth to emerge \citep{cavalazzietal21}.

A precise estimation of the eccentricity evolution requires a more comprehensive treatment of the coupled orbital and rotational evolution, as discussed by \citet{hut81, eggleton-etal-98, mardling-lin-02}, who showed that tidal dissipation, angular-momentum exchange, and resonance locking can significantly modify the damping rates derived from simplified constant-phase-lag models.
This will be the subject of a future study.

\subsection{A note on observability}
\label{sec:disc:obs}

Rogue planets can be detected mainly by direct imaging \citep{zapateroosioroetal00} or by their gravitational microlensing effect \citep{mrozetal18, mrozetal19, mrozetal20}.
Concerning the latter method, it cannot be ruled out with absolute certainty that rogue planets are ordinary planets orbiting their stars in very wide orbits.

The candidate detection by \citet{limbachetal21} of a moon transiting a rogue planet, though not yet confirmed, points to the necessity of using the latest generation of instruments, such as the James Webb Space Telescope, to observe exomoons occulting around rogue planets.
This assumption is further supported by the fact that moons the size of Titan or Ganymede can cause a brightness decrease of up to 2\% as they pass by their planet \citep{petersturner13}.
This method favours the detection of rogue systems with a high mass ratio because the transit depth is proportional to the size of the moons.

Observing the gravitational microlensing effect of rogue planet moons requires a lucky lensing setup \citep{mrozetal19} and the construction of new multi-lensing models for the evaluation \citep{hanhan02}.
So far, two microlensing events were detected where it cannot be excluded that the lensing object was a rogue planet-moon system \citep{bennettetal14, hwangetal18}.
Note that, however, a distant star-planet and a nearby planet-moon lensing system produce the same signal, and in the above observations, an assumption of a distant star-planet system provided the best fit to the data.
These measurements, however, demonstrate that rogue moons are detectable even with our current technology and especially with the next generation of instruments: 
the Euclid and Nancy Grace Roman Space Telescope sky surveys.

The brightening due to microlensing is proportional to the mass of the planet.
Furthermore, the lower the rogue systems' peculiar velocity, the longer the microlensing effect is observed.
This is supported by that the moons considered to be urable in Sect.~\ref{sec:disc:urability} are orbiting far from their stars, meaning that they will have small peculiar velocities, as discussed in Sect.~\ref{sec:results:base} and in Fig.~\ref{fig:vpec}.

\subsection{Caveats of the model}
\label{sec:disc:caveats}

We use a homologous expansion model that assumes spherical symmetry.
Perturbations arising due to an asymmetric envelope ejection can be different:
asymmetric kicks can alter the planetary and stellar peculiar velocities and orbital elements significantly compared to those reported in this study \citep{parriott-alcock, namouni, namouni-zhou}.
However, developing such a model is beyond the scope of the current paper.

Due to the nature of the one-dimensional approximation in the homologous expansion model, interactions between the planet-moon system and the ejected envelope were also not investigated.
These interactions could disrupt the spherical symmetry and also cause a significant mass loss of giant planets' gas envelopes \citep{debes-sigurdsson}.
This mass loss could introduce additional perturbations in the orbital elements of the moon and could further increase its eccentricity.
However, the tidal heating of the moon would decrease due to the planetary mass loss.
A thorough investigation would require the establishment of a mass loss and accretion model for the planetary atmosphere, which is postponed to a future study.

Whether eccentric planets remain bound to the neutron star depends strongly on their pre-explosion true anomaly. 
In the parameter space explored here, a bound post-explosion orbit requires $f_{\mathrm{p}}^{(0)} = 180^\circ$ and an eccentricity of $e_{\mathrm{p}}^{(0)} \geq 0.6$. 
Precisely determining the range of true anomalies that allow the planet to remain bound would require much finer sampling, similar to the approach used in \citet{regalyetal22}.

The temperature of exomoons is not determined only by stellar irradiation and tidal heating, as explained in Sect.~\ref{sec:disc:urability}.
In a more comprehensive model, one should take into account the reflected light from the planet, the planet's thermal emission, the residual heat from the moon's formation, and the heat generated by the decay of radioactive isotopes in the moon.
Although the starlight reflected by a rogue planet is irrelevant and its thermal radiation is significantly reduced over time, the internal heat sources of the moon can keep its inner layers warm, even when it is a rogue body \citep{kasting}.

The analytical approximations used to model the tidal heating and the timescale of eccentricity damping ignore the changes in the orbital elements of the moon.
Tidal evolution depends significantly on the relation between the rotation and orbital rates of the planet and the moon.
Moons that orbit in a retrograde manner or that are prograde but orbit below the synchron radius are subject to a decrease in semi-major axis.
On the other hand, the semi-major axis of prograde moons outside of the synchron radius will increase over time \citep{murraydermott}.
To model the effects described above, a more sophisticated tidal model should be developed, as proposed by e.g. \citet{goldreichpeale, ferrazmello12, boueefriomsky19}.

The effective temperature of the moon is calculated using the Stefan-Boltzmann law, similar to the work of \citet{tokadjianpiro23}, given the tidal heating power (see Sect.~\ref{sec:disc:urability}).
This assumes that tidal effects heat the entire volume of the moon.
In realistic systems, however, the distribution of heat depends heavily on the composition and internal structure of the moon.
Therefore, in the future, it will be necessary to model moons that consist of several layers, such as a core, mantle, and ice crust, taking into account thermal conduction between these layers and their different densities.

\section{Conclusions}
\label{sec:conclusions}

To address the question of whether life could exist on other celestial bodies, it is essential to identify the minimal physical, chemical, and energetic conditions required for life to originate.
Bodies that meet these criteria are referred to as urable.
Exomoons are promising candidates in the search for urable environments, as they can be heated not only by stellar irradiation but also by tidal forces.
In our own Solar System, moons such as Europa, Enceladus, Mimas, and Miranda exhibit significant tidal heating, which can melt internal ice layers and sustain subsurface oceans tens of kilometres deep beneath their icy crust.
If similar tidal processes are active in moons orbiting rogue planets, such bodies may remain urable over long timescales, even in the absence of a host star.

In this study, we have investigated the dynamical stability and tidal heating of planet-moon systems whose massive host star undergoes significant mass loss during an SN~II explosion.
We modelled the hierarchical triple system using an 8th-order Runge--Kutta integrator, incorporating a time-varying central mass and a homologous expansion model for the stellar envelope to describe the dynamical effects of the SN~II explosion.
We monitored the perturbations of the orbital elements of the planets and moons and the peculiar velocity of the bodies leaving the system across four different model sets:
(i) circular pre-explosion orbits for both the planet and the moon; (ii) eccentric pre-explosion planetary orbits; (iii) eccentric pre-explosion orbits for the moons; and (iv) two moons in mean-motion resonance around the planet.
To determine whether the moons of rogue planets can be urable, we have analytically calculated the specific tidal heating power and the effective temperature of the moons pre- and post-explosion.
The main results obtained from the analysis of the N-body simulations and tidal heating calculations are summarised below.

\begin{enumerate}

    \item In the parameter space studied, planets are expelled from their host system post-explosion and continue their life as rogue planets.
    The only exception to this is planets that are at the apocentre of a highly eccentric orbit ($e_{\mathrm{p}}^{(0)}\geq0.6$) at the moment of explosion.
    In all cases, the moons remain bound to their planets post-explosion.
    The change in the semi-major axis of the moons is negligible in all cases, and is found to be no greater than 0.2\%.
    The peculiar velocity of the rogue planet-moon systems is found to be $6-32~\mathrm{km~s^{-1}}$.

    \item If both the planet and the moon are on circular orbits pre-explosion, the moons acquire an eccentricity of $\lesssim7.3\times10^{-4}$ post-explosion. 
    Their eccentricity is independent of the envelope ejection velocity, but decreases with the semi-major axis of the planet and increases with planet-moon distance.
    The pre-explosion true anomaly and inclination of the moons have no effect on their post-explosion eccentricity and semi-major axis.
    
    \item Regarding non-circular orbits, 
    (a) if the planets orbited with an eccentricity of 0.1-0.8 pre-explosion, the post-explosion eccentricities of the moons were found to be $\lesssim3\times10^{-3}$; 
    (b) if the moons had eccentric orbits pre-explosion ($e_{\mathrm{m}}^{(0)}\in [10^{-4}-10^{-1}]$), their eccentricities remain effectively unchanged afterward, and the post-explosion change in eccentricity is found to be $\lesssim 3\times10^{-4}$; and
    (c) moons that were in resonant configurations pre-explosion remain in resonance afterward, while their eccentricities were found to be in the range of approximately $10^{-5}$ to $10^{-2}$.

    \item The specific tidal heating power on the moons reaches at least one-tenth of that observed on Europa and Enceladus when their post-explosion eccentricity is $e_\mathrm{m}^{(1)} \gtrsim [10^{-5} - 10^{-3}]$, depending on their density and tidal dissipation factor.
    Moons meeting this criterion are considered urable following the SN~II explosion.
    Urable moons are found across all investigated combinations (see Table~\ref{tab:params}) of planetary semi-major axis, planet–moon distance, and planetary mass.

    \item The urability of moons around rogue planets also requires that their water reservoirs do not boil pre-explosion.
    Within the parameter space explored, this condition is met for planet–moon systems with $a_{\mathrm{p}}^{(0)} \gtrsim 383$au, $M_{\mathrm{p}} \geq 0.1~M_{\mathrm{Jup}}$, $a_{\mathrm{m}}^{(1)} \in [4;36]~R_{\mathrm{p}}$, and $e_{\mathrm{m}}^{(1)} \in [0.1-0.9]$.
    
    \item The eccentricity dampening timescale of the moons in the parameter space studied is $10^{4}-10^{12}$~years.
    Moons orbiting farther than $\simeq 10$ planetary radii retain their eccentricity -- and thus, urability -- longer than the age of the Solar System.
\end{enumerate}

In summary, the eccentricity excitation of exomoons resulting from the mass loss during an SN~II explosion is generally limited. 
For moons initially on circular orbits, achieving eccentricities comparable to those of Europa ($5\times10^{-3}$) or Enceladus ($9\times10^{-3}$) requires the host planet to have been near pericentre in a highly eccentric pre-explosion orbit. 
In contrast, moons that were already eccentric before the explosion (whether individually or as members of a resonant pair) retain their eccentricities post-explosion. 
Our findings suggest that such eccentric moons, once expelled into interstellar space with their host planet, may remain urable for billions of years, sustained by tidal heating alone.
Importantly, this reasoning extends beyond supernova scenarios: the potential urability of exomoons applies equally to eccentric rogue planet–moon systems produced by early dynamical instabilities or stellar flybys.

\begin{acknowledgements}
We thank the anonymous referee for their helpful comments, which significantly improved the quality of this paper.
V. F. acknowledges financial support from the undergraduate research assistant program of the Konkoly Observatory and the EKÖP-24-
2-I-ELTE-64 University Research Scholarship Program of the Ministry for Culture and Innovation
from the source of the National Research, Development and Innovation Fund.
The authors express their gratitude to Tamás Csizmadia for insightful discussions on the origins of life and the potential urability of moons orbiting rogue planets.
\end{acknowledgements}

\section*{Data availability}
The data underlying this article can be shared at a reasonable request to the corresponding author.

\begin{appendix}

\section{Homologous envelope expansion model}
\label{sec:apx:homologous}

\begin{figure*}[ht!]
    \centering
    \includegraphics[width=0.9\linewidth]{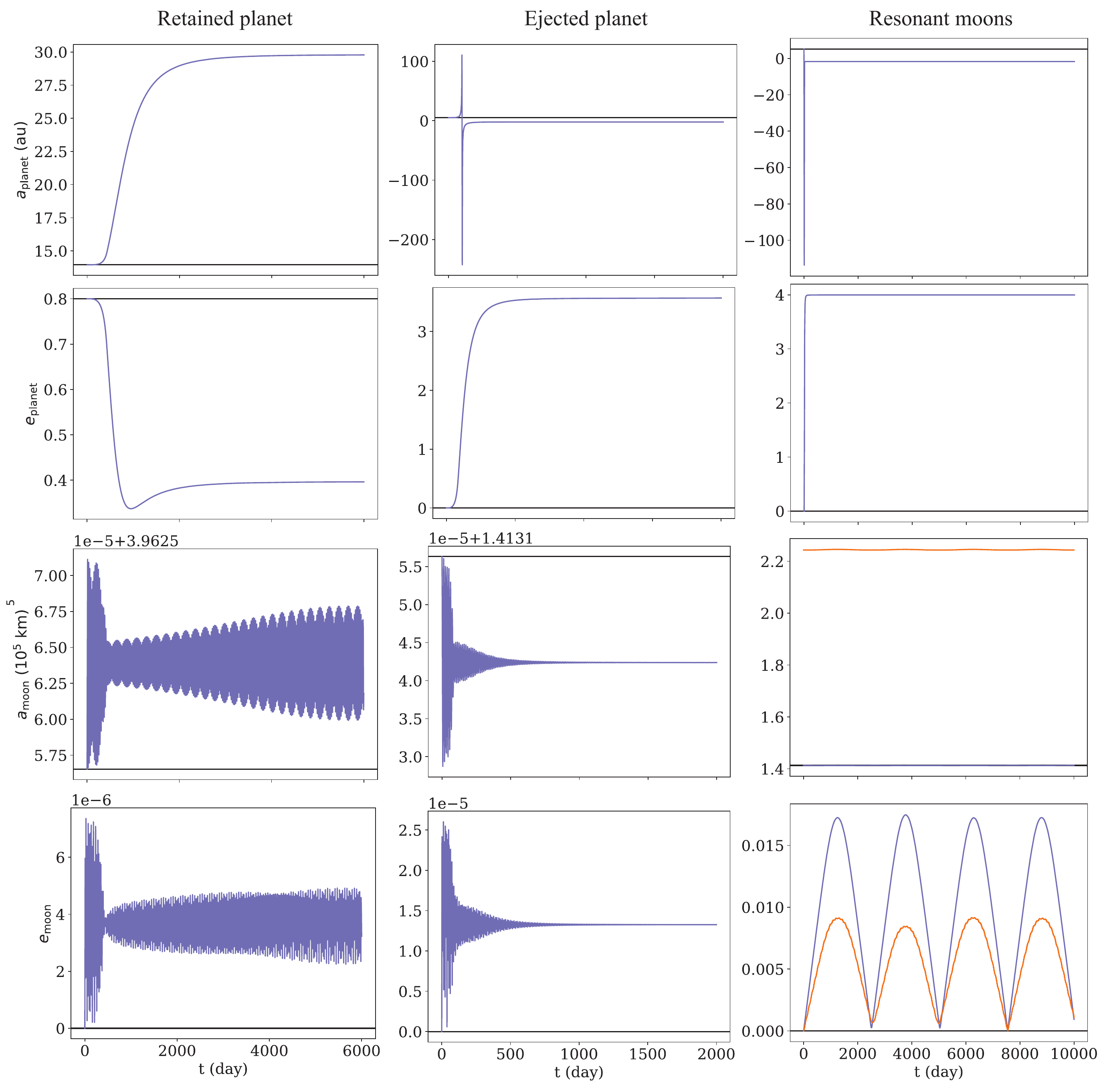}
    \caption{
    Temporal evolution of the orbital elements of planets and moons in three representative simulations. In the resonant case, the two moons' orbital elements are depicted with different colours.
    }
    \label{fig:orbelems}
\end{figure*}

During the SN~II explosion, the central star undergoes significant mass loss, which perturbs the orbit of the companions: a planet and its moon.
The equations of motion of the bodies, 
\begin{equation}
    \ddot{\textbf{r}}_\mathrm{s}= - M_\mathrm{p} \ G \ \frac{\textbf{r}_\mathrm{s}-\textbf{r}_\mathrm{p}}{|\textbf{r}_\mathrm{s}-\textbf{r}_\mathrm{p}|^{3}}- M_\mathrm{m} \ G \ \frac{\textbf{r}_\mathrm{s}-\textbf{r}_\mathrm{m}}{|\textbf{r}_\mathrm{s}-\textbf{r}_\mathrm{m}|^{3}},
    \label{eq:s}
\end{equation}
\begin{equation}
    \ddot{\textbf{r}}_\mathrm{p}= - M_\mathrm{in,p} \ G \ \frac{\textbf{r}_\mathrm{p}-\textbf{r}_\mathrm{s}}{|\textbf{r}_\mathrm{p}-\textbf{r}_\mathrm{s}|^{3}}- M_\mathrm{m} \ G \ \frac{\textbf{r}_\mathrm{p}-\textbf{r}_\mathrm{m}}{|\textbf{r}_\mathrm{p}-\textbf{r}_\mathrm{m}|^{3}},
    \label{eq:p}
\end{equation}
\begin{equation}
    \ddot{\textbf{r}}_\mathrm{m}= - M_\mathrm{in,m} \ G \ \frac{\textbf{r}_\mathrm{m}-\textbf{r}_\mathrm{s}}{|\textbf{r}_\mathrm{m}-\textbf{r}_\mathrm{s}|^{3}}- M_\mathrm{p} \ G \ \frac{\textbf{r}_\mathrm{m}-\textbf{r}_\mathrm{p}}{|\textbf{r}_\mathrm{m}-\textbf{r}_\mathrm{p}|^{3}},
    \label{eq:m}
\end{equation}
are solved numerically in three dimensions, using an 8th-order Runge--Kutta method.
In the above equations, $G$ is the gravitational constant,  $\textbf{r}_\mathrm{s}$, $\textbf{r}_\mathrm{p}$, and $\textbf{r}_\mathrm{m}$ are the position vectors of the star, planet, and moon in the inertial frame, respectively.
Integration time is 100,000 days ($\approx$270 years), which allows the monitored values (semi-major axis $a$, eccentricity $e$, and velocities $v$) of the star, planet, and moon to saturate by the end of the simulations.
Simulating a fixed-mass three-body problem with a high-mass ($10~M_{\mathrm{Jup}}$), high-eccentricity ($0.8$) planet, the relative error in total energy is on the order of $10^{-12}$ by the end of integration.

A planet or moon is considered to be bound if its eccentricity remains under unity.
The mass of the planet, $M_\mathrm{p}$, and that of the moon, $M_\mathrm{m}$, are constants.
The stellar mass inside the planet's and the moon's orbits are $M_{\mathrm{in,p}}$ and $M_{\mathrm{in,m}}$, respectively.

The conversion of the orbital elements to Cartesian coordinates and vice versa is done as follows.
From Keplerian orbital elements to Cartesian coordinates:
\begin{align}
E &= 2 \arctan\!\left( 
      \sqrt{\frac{1 - e}{1 + e}} \, \tan\! \left(\frac{f}{2}\right)
    \right), \\[1em]
d &= a \left( 1 - e \cos E \right), \\
\mathbf{r}_\mathrm{o} &= d 
  \begin{bmatrix}
    \cos f \\
    \sin f \\
    0
  \end{bmatrix},
~~~~\mathbf{v}_\mathrm{o} = 
  \frac{\sqrt{\mu a}}{d}
  \begin{bmatrix}
    -\sin E \\
    \sqrt{1 - e^2}\, \cos E \\
    0
  \end{bmatrix},
\end{align}
where $E$ is the eccentric anomaly, $\mu$ the gravitational parameter of the system, and $a$, $e$, and $f$ are the semi-major axis, eccentricity, and true anomaly, respectively.
Transforming to the inertial reference frame using the inclination $i$ and assuming $\omega=\Omega=0$, the position and velocity vectors are calculated as:
\begin{equation}
\mathbf{r} =
\begin{bmatrix}
r_\mathrm{o,x} \\[0.4em]
r_\mathrm{o,y}\cos i \\[0.4em]
r_\mathrm{o,y}\sin i
\end{bmatrix},
~~~~\mathbf{v} =
\begin{bmatrix}
v_\mathrm{o,x} \\[0.4em]
v_\mathrm{o,y}\cos i \\[0.4em]
v_\mathrm{o,y}\sin i
\end{bmatrix}.
\end{equation}
From the Cartesian coordinates to the Keplerian orbital elements, the conversion happens as
\begin{equation}
\begin{aligned}
    \vec{c} &= \vec{r} \times \vec{v} \\
    h &= \frac{1}{2}v^2-\frac{\mu}{r} \\
    e &= \sqrt{1+2h\left(\frac{c}{\mu}\right)^2} \\
    a &= -\frac{\mu}{2h} \\
    i &= \arctan2( \sqrt{(yv_z-zv_y)^2+(-xv_z+zv_x)^2} , (xv_y-yv_x) ),
\end{aligned}
\end{equation}
where $\vec{c}$ is the specific angular momentum vector, $c$ its magnitude, $h$ the specific energy of the system, while $x,y,z$ and $v_x, v_y, v_z$ are the components of the position and velocity vectors of the companion.

At the first step of the integration, both of the above masses are equal to the pre-explosion stellar mass, $M_{\mathrm{ej}}+M_{\mathrm{n}}$, where $M_{\mathrm{ej}}$ is the mass of the envelope, and $M_{\mathrm{n}}$ is that of the remnant neutron star. 
Later, they change according to the homologous expansion model \citep{arnett80, vinko04, bw17, regalyetal22, frohlichetal23, regalyetal25}.

Homologous expansion means that the velocity of the ejected layers is a linear function of distance from the SN, and that the velocity of the outermost layer and the density profile of the ejected material are time-invariant.
Three cases are distinguished when applying the model:
1) all mass resides within the companion's orbit, $M_{\mathrm{in}}=M_{\mathrm{ej}}+M_{\mathrm{n}}$;
2) the co-moving coordinate $x$ of the companion is larger than that of the core, $x_\mathrm{comp}>x_\mathrm{c}$), thus
\begin{equation}
        M_\mathrm{in} = M_\mathrm{n} + M_\mathrm{c} \left [ 1+ \frac{3}{n-3} \left ( 1- \left ( \frac{x_\mathrm{comp}}{x_\mathrm{c}} \right )^{3-n} \right ) \right ];
    \label{eq:min_xp>xc}
\end{equation}
3) the co-moving distance is smaller than that of the core, $x_\mathrm{comp}<x_\mathrm{c}$, thus
\begin{equation}
    M_\mathrm{in} = M_\mathrm{n} + M_\mathrm{c} \left ( \frac{x_\mathrm{comp}}{x_\mathrm{c}} \right )^3.
    \label{eq:min_xp<xc}
\end{equation}
The stellar envelope contains a constant-density core extending up to a fractional radius of $x_{\mathrm{c}} = r_{\mathrm{c}} / R_{0}$, where $r_{\mathrm{c}}$ is the initial radius of the core and $R_{0}$ that of the progenitor.
In the above equations, $M_{\mathrm{c}}$ is the mass of the core of the envelope.
The density of the envelope follows a power law with an exponent $n=7$.

Figure \ref{fig:orbelems} shows the temporal evolution of the orbital elements of both the planet and the moon in three representative cases.
The left panels show the case where both the planet and the moon stay on bound orbits post-explosion.
The middle panels show a scenario where the planet leaves the system and continues as a rogue object, yet the moon still stays bound to it.
The rightmost panels show a system where two moons are placed in resonance pre-explosion, and where the moon stays bound to a rogue planet post-explosion.

\section{Analytical tidal heating models}
\label{sec:apx:tidal}

For moons in eccentric orbits, time-varying forces cause part of the moon's orbital energy to contribute to heating the moon (see, e.g., \citealp{kaula64, pealeetal79, ogilvielin04}).
Here, we present analytical models describing the magnitude of tidal heating, which can be used to calculate the effective temperature of moons of rogue planets.
For a complete treatment of the coupled evolution of semi-major axis, eccentricity, and spin within the equilibrium-tide framework, see \citet{hut81, eggleton-etal-98, mardling-lin-02}.

We investigate tidal heating by implementing two different models.
One is the so-called constant phase lag (CPL) model \citep{macdonald64, efroimskymakarov13}.
In the CPL model, the tidal friction causes the tidal bulge to overtake or lag behind the moon, while the angle of divergence is constant.
Note that this model neglects the higher-order eccentricity contributions.
In the CPL model, the instantaneous power of tidal heating can be written as
\begin{equation}
    L_{\mathrm{tidal,CPL}}= \frac{21}{2} G \frac{k_{2,\mathrm{m}}}{Q_{\mathrm{m}}} M_{\mathrm{p}}^{5/2} R_{\mathrm{m}}^{5} \frac{e^2}{a^{15/2}}.
\end{equation}
In the above equation, $k_{2,\mathrm{m}}$ is the second-order Love number of the moon, which measures its deformability by tidal forces, while
$Q_{\mathrm{m}}$ is the moon's tidal dissipation function, which tells us how efficiently the moon dissipates orbital energy through tidal forces.
$M_{\mathrm{p}}$ is the planet's mass, $R_{\mathrm{m}}$ is the moon's physical radius, and $a$ and $e$ are the moon's semi-major axis and eccentricity, respectively.
Note that an increase in the semi-major axis or a decrease in the eccentricity both decrease tidal heating power.

The other tidal heating model implemented in this study is the constant time lag (CTL) model, where the tidal bulge follows the motion of the celestial bodies with a temporal delay $\tau_{\mathrm{m}}$.
According to \citet{hellerbarnes13}, the tidal luminosity in the CTL model is
\begin{equation}
    L_{\mathrm{tidal,CTL}}= \frac{3 G^2 k_{2,\mathrm{m}} M_{\mathrm{p}}^2 (M_{\mathrm{p}}+M_{\mathrm{m}}) R_{\mathrm{m}}^5 a^{-9}}{(1-e^2)^{15/2}} \left ( f_1 - \frac{f_2^2}{f_5} \right ) \tau_{\mathrm{m}},
\end{equation}
\begin{equation}
    \begin{aligned}
        f_1 &= 1 + \frac{31}{2} e^2 + \frac{255}{8} e^4 + \frac{188}{16} e^6 + \frac{25}{64} e^8\\
        f_2 &= 1 + \frac{15}{2} e^2 + \frac{45}{8} e^4 + \frac{5}{16} e^6 \\
        f_5 &= 1 + 3 e^2 + \frac{3}{8} e^4 .
    \end{aligned}
\end{equation}

\begin{figure}[ht!]
    \centering
    \includegraphics[width=0.9\columnwidth]{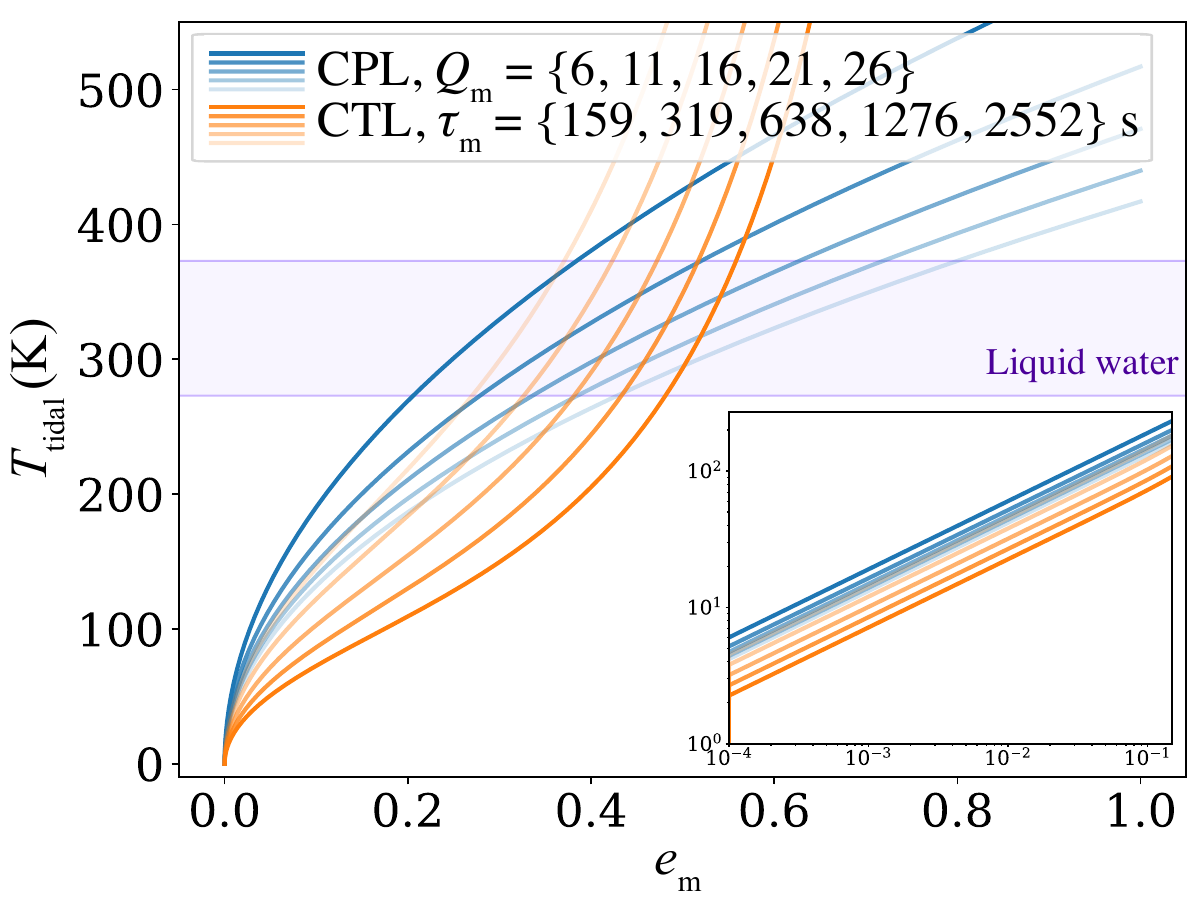}
    \caption{
    Comparison of the CPL (blue lines, under different dissipation functions) and CTL (orange lines, under different time lags) tidal heating models assuming the physical parameters and semi-major axis of the Jupiter-Europa system. 
    The effective temperature of the moon is derived from tidal heating as a function of the moon's assumed eccentricity.
    The purple band shows the region where water can exist in a liquid form.
    }
    \label{fig:heating-models}
\end{figure}

We compare the tidal temperature in the CPL and CTL models.
The tidal temperature is calculated according to the Stefan--Boltzmann law, which implies that $L_{\mathrm{tidal}}=A_{\mathrm{m}} \sigma T_{\mathrm{m}}^4$, if there is no other heat source.
Thus, the effective temperature of the moon due to tidal heating is
\begin{equation}
    T_{\mathrm{m}}=\left( \frac{L_{\mathrm{tidal}}}{4\pi \sigma R_{\mathrm{m}}^{2}}\right)^{1/4}.
\end{equation}
Figure \ref{fig:heating-models}.~ shows the comparison of the two models assuming the physical parameters and semi-major axis of the Jupiter-Europa system at different eccentricities. 
It can be seen that the results of the two models are qualitatively the same at low eccentricities ($e_{\mathrm{m}}\lesssim0.1$).
At properly chosen parameters of $Q_{\mathrm{m}}=26$ and $\tau_{\mathrm{m}}=2552$, the two models are indistinguishable.
At higher eccentricities, however, the models show a qualitative difference, with the CTL model yielding higher effective temperatures.
Note that in our case, the post-explosion eccentricity of the moons of rogue planets is never larger than 0.1, so either of the models can be applied.
In this study, we use the CPL model.

\end{appendix}


\begin{thebibliography}{200}

\bibitem[{\'A}vila et al.(2021)]{avilaetal21} {\'A}vila, P.~J., Grassi, T., Bovino, S., et al.\ 2021, International Journal of Astrobiology, 20, 300. 
\bibitem[Abbot \& Switzer(2011)]{abbottswitzer11} Abbot, D.~S. \& Switzer, E.~R.\ 2011, \apjl, 735, L27. 
\bibitem[Andrade(1910)]{mutilde} Andrade, E.~N.~D.~C.\ 1910, Proceedings of the Royal Society of London Series A, 84, 1
\bibitem[Arnett(1980)]{arnett80} Arnett, W.~D.\ 1980, \apj, 237, 541. 
\bibitem[Bachelet et al.(2022)]{bacheletetal22} Bachelet, E., Specht, D., Penny, M., et al.\ 2022, \aap, 664, A136. 
\bibitem[Baross \& Hoffman(1985)]{barosshoffman85} Baross, J.~A. \& Hoffman, S.~E.\ 1985, Origins of Life, 15, 327. 
\bibitem[Bear et al.(2011)]{bear-etal} Bear, E., Soker, N., \& Harpaz, A.\ 2011, \apjl, 733, L44.
\bibitem[Bennett et al.(2014)]{bennettetal14} Bennett, D.~P., Batista, V., Bond, I.~A., et al.\ 2014, \apj, 785, 155. 
\bibitem[Bou{\'e} \& Efroimsky(2019)]{boueefriomsky19} Bou{\'e}, G. \& Efroimsky, M.\ 2019, Celestial Mechanics and Dynamical Astronomy, 131, 30. 
\bibitem[Boyd \& Whitworth(2005)]{boydwhitworth05} Boyd, D.~F.~A. \& Whitworth, A.~P.\ 2005, \aap, 430, 1059. 
\bibitem[Branch \& Wheeler(2017)]{bw17} Branch, D. \& Wheeler, J.~C.\ 2017, Supernova Explosions: Astronomy and Astrophysics Library, ISBN 978-3-662-55052-6. Springer-Verlag GmbH Germany, 2017. 
\bibitem[Burns(1976)]{burns} Burns, J.~A.\ 1976, American Journal of Physics, 44, 944.
\bibitem[Canup \& Ward(2002)]{canupward02} Canup, R.~M. \& Ward, W.~R.\ 2002, \aj, 124, 3404.
\bibitem[Canup \& Ward(2006)]{canupward06} Canup, R.~M. \& Ward, W.~R.\ 2006, \nat, 441, 834.
\bibitem[Canup \& Ward(2009)]{canupward09} Canup, R.~M. \& Ward, W.~R.\ 2009, Europa, 59. 
\bibitem[Carr et al.(1998)]{carretal98} Carr, M.~H., Belton, M.~J.~S., Chapman, C.~R., et al.\ 1998, \nat, 391, 363. 
\bibitem[Cavalazzi et al.(2021)]{cavalazzietal21} Cavalazzi, B., Hickman-Lewis, K., Brack, A., et al.\ 2021, Prebiotic Chemistry and the Origin of Life, 227. 
\bibitem[Chamandy et al.(2021)]{chamandyetal} Chamandy, L., Blackman, E.~G., Nordhaus, J., et al.\ 2021, \mnras, 502, L110.]
\bibitem[Clanton \& Gaudi(2016)]{clantongaudi16} Clanton, C. \& Gaudi, B.~S.\ 2016, \apj, 819, 125. 
\bibitem[Costa et al.(2019a)]{costaetal19a} Costa, G., Girardi, L., Bressan, A., et al.\ 2019, \aap, 631, A128. 
\bibitem[Costa et al.(2019b)]{costaetal19b} Costa, G., Girardi, L., Bressan, A., et al.\ 2019, \mnras, 485, 4641. 
\bibitem[Crida \& Charnoz(2012)]{cridacharnoz12} Crida, A. \& Charnoz, S.\ 2012, Science, 338, 1196. 
\bibitem[Deamer et al.(2022)]{deameretal22} Deamer, D., Cary, F., \& Damer, B.\ 2022, Astrobiology, 22, 889. 
\bibitem[Debes \& Sigurdsson(2007)]{debessigurdsson07} Debes, J.~H. \& Sigurdsson, S.\ 2007, \apjl, 668, L167. 
\bibitem[Debes \& Sigurdsson(2002)]{debes-sigurdsson} Debes, J.~H. \& Sigurdsson, S.\ 2002, \apj, 572, 556.
\bibitem[Debras et al.(2021)]{debrasetal21} Debras, F., Baruteau, C., \& Donati, J.-F.\ 2021, \mnras, 500, 2, 1621.
\bibitem[Dobos \& Turner(2015)]{dobosturner15} Dobos, V. \& Turner, E.~L.\ 2015, \apj, 804, 1, 41. 
\bibitem[Dobos et al.(2021)]{dobosetal21} Dobos, V., Charnoz, S., P{\'a}l, A., et al.\ 2021, \pasp, 133, 094401.
\bibitem[Doolin \& Blundell(2011)]{doolinblundell11} Doolin, S. \& Blundell, K.~M.\ 2011, \mnras, 418, 2656. 
\bibitem[Efroimsky \& Makarov(2013)]{efroimskymakarov13} Efroimsky, M. \& Makarov, V.~V.\ 2013, \apj, 764, 26. 
\bibitem[Eggleton et al.(1998)]{eggleton-etal-98} Eggleton, P.~P., Kiseleva, L.~G., \& Hut, P.\ 1998, \apj, 499, 2, 853.
\bibitem[Estrada et al.(2009)]{estradaetal09} Estrada, P.~R., Mosqueira, I., Lissauer, J.~J., et al.\ 2009, Europa, 27. 
\bibitem[Fagginger Auer \& Portegies Zwart(2022)]{faggingerauer-portegieszwart-2021} Fagginger Auer, F. \& Portegies Zwart, S.\ 2022, SciPost Astronomy, 2, 002.
\bibitem[Ferraz-Mello(2013)]{ferrazmello12} Ferraz-Mello, S.\ 2013, Celestial Mechanics and Dynamical Astronomy, 116, 109.
\bibitem[Ford \& Rasio(2008)]{fordrasio08} Ford, E.~B. \& Rasio, F.~A.\ 2008, \apj, 686, 621.
\bibitem[Fox \& Wiegert(2021)]{foxwiegert21} Fox, C. \& Wiegert, P.\ 2021, \mnras, 501, 2378. 
\bibitem[Fr{\"o}hlich et al.(2023)]{frohlichetal23} Fr{\"o}hlich, V., Reg{\'a}ly, Z., \& Vink{\'o}, J.\ 2023, \mnras, 523, 4957. 
\bibitem[Fryer(1999)]{fryer} Fryer, C.~L.\ 1999, \apj, 522, 413.
\bibitem[Gifford et al.(2024)]{giffordetal24} Gifford, D. R., Bhattacharyya, A., Geim, A., Marshall, E., Krašovec, R., \& Knight, C. G. Microbiology, 170(4), 001452. 
\bibitem[Goldreich \& Peale(1966)]{goldreichpeale} Goldreich, P. \& Peale, S.\ 1966, \aj, 71, 425. 
\bibitem[Greenberg(2005)]{eurinnertemp} Greenberg, R.\ 2005, 380 p.  ISBN 3540224505.  Chichester, UK: Springer, 2005
\bibitem[Groh et al.(2013)]{grohetal13} Groh, J.~H., Meynet, G., Georgy, C., et al.\ 2013, \aap, 558, A131. 
\bibitem[Guillochon et al.(2011)]{guillochanetal11} Guillochon, J., Ramirez-Ruiz, E., \& Lin, D.\ 2011, \apj, 732, 74. 
\bibitem[Gyld{\'e}n(1884)]{gylden} Gyld{\'e}n, H.\ 1884, Astronomische Nachrichten, 109, 1.
\bibitem[Hadjidemetriou(1963)]{hadjidemetriou-1963} Hadjidemetriou, J.~D.\ 1963, \icarus, 2, 440. 
\bibitem[Hadjidemetriou(1966a)]{hadjidemetriou-1966-a} Hadjidemetriou, J.~D.\ 1966a, \icarus, 5, 34.
\bibitem[Hadjidemetriou(1966b)]{hadjidemetriou-1966-b} Hadjidemetriou, J.~D.\ 1966b, \zap, 63, 116
\bibitem[Hamuy \& Pinto(2002)]{hamuy-pinto} Hamuy, M. \& Pinto, P.~A.\ 2002, \apjl, 566, L63. 
\bibitem[Han \& Han(2002)]{hanhan02} Han, C. \& Han, W.\ 2002, \apj, 580, 490. 
\bibitem[Han et al.(2014)]{hanetal14} Han, E., Wang, S.~X., Wright, J.~T., et al.\ 2014, \pasp, 126, 827.
\bibitem[Heller \& Armstrong(2014)]{hellerarmstrong14} Heller, R. \& Armstrong, J.\ 2014, Astrobiology, 14, 1, 50.
\bibitem[Heller \& Barnes(2013)]{hellerbarnes13} Heller, R. \& Barnes, R.\ 2013, Astrobiology, 13, 18. 
\bibitem[Heller \& Barnes(2015)]{hellerbarnes15} Heller, R. \& Barnes, R.\ 2015, International Journal of Astrobiology, 14, 335.
\bibitem[Heller et al.(2019)]{helleretal19} Heller, R., Rodenbeck, K., \& Bruno, G.\ 2019, \aap, 624, A95. 
\bibitem[Heller(2018)]{heller18} Heller, R.\ 2018, Handbook of Exoplanets, 35. 
\bibitem[Henderson \& Shvartzvald(2016)]{henderson16} Henderson, C.~B. \& Shvartzvald, Y.\ 2016, \aj, 152, 96. 
\bibitem[Holman \& Wiegert(1999)]{holmanwiegert99} Holman, M.~J. \& Wiegert, P.~A.\ 1999, \aj, 117, 621. 
\bibitem[Hong et al.(2018)]{hongetal18} Hong, Y.-C., Raymond, S.~N., Nicholson, P.~D., et al.\ 2018, \apj, 852, 85. 
\bibitem[Howett et al.(2010)]{encalbedo} Howett, C.~J.~A., Spencer, J.~R., Pearl, J., et al.\ 2010, \icarus, 206, 573.
\bibitem[Hurley et al.(2000)]{hurley-etal} Hurley, J.~R., Pols, O.~R., \& Tout, C.~A.\ 2000, \mnras, 315, 543.
\bibitem[Hut(1981)]{hut81} Hut, P.\ 1981, \aap, 99, 126
\bibitem[Hwang et al.(2018)]{hwangetal18} Hwang, K.-H., Udalski, A., Bond, I.~A., et al.\ 2018, \aj, 155, 259. 
\bibitem[Irani et al.(2024)]{Irani2024ApJ...970...96I} Irani, I., Morag, J., Gal-Yam, A., et al.\ 2024, \apj, The Early Ultraviolet Light Curves of Type II Supernovae and the Radii of Their Progenitor Stars, 970, 1, 96. 
\bibitem[Irwin \& Schulze-Makuch(2020)]{irwinschulze20} Irwin, L.~N. \& Schulze-Makuch, D.\ 2020, Universe, 6, 130.
\bibitem[J{\"a}ger \& Szab{\'o}(2021)]{jaegerszabo21} J{\"a}ger, Z. \& Szab{\'o}, G.~M.\ 2021, \mnras, 508, 5524. 
\bibitem[Jackson et al.(2008)]{jacksonetal08} Jackson, B., Barnes, R., \& Greenberg, R.\ 2008, \mnras, 391, 237. 
\bibitem[Jacobson(2022)]{encdensity} Jacobson, R.~A.\ 2022, \aj, 164, 199.
\bibitem[K{\'a}lm{\'a}n et al.(2023)]{kalmanetal23} K{\'a}lm{\'a}n, S., Szab{\'o}, G.~M., \& Csizmadia, S.\ 2023, \aap, 675, A107. 
\bibitem[Kaib et al.(2013)]{kaibetal13} Kaib, N.~A., Raymond, S.~N., \& Duncan, M.\ 2013, \nat, 493, 381. 
\bibitem[Karve \& Wagner(2022)]{karvewagner22}Karve, S., Wagner, A. \nat ~Commun. 13, 5904 (2022) 
\bibitem[Kasting et al.(1993)]{kasting} Kasting, J.~F., Whitmire, D.~P., \& Reynolds, R.~T.\ 1993, \icarus, 101, 108.
\bibitem[Kaula(1964)]{kaula64} Kaula, W.~M.\ 1964, Reviews of Geophysics and Space Physics, 2, 661.
\bibitem[Kennedy \& Kenyon(2008)]{kennedykenyon08} Kennedy, G.~M. \& Kenyon, S.~J.\ 2008, \apj, 673, 502.
\bibitem[Kipping et al.(2009)]{kippingetal09} Kipping, D.~M., Fossey, S.~J., \& Campanella, G.\ 2009, \mnras, 400, 398. 
\bibitem[Kipping et al.(2022)]{kippingetal22} Kipping, D., Bryson, S., Burke, C., et al.\ 2022, Nature Astronomy, 6, 367. 
\bibitem[Kipping(2009)]{kipping09ttv} Kipping, D.~M.\ 2009, \mnras, 396, 1797. 
\bibitem[Kipping(2011)]{kipping11} Kipping, D.~M.\ 2011, \mnras, 416, 689. 
\bibitem[Kipping(2020)]{kipping20} Kipping, D.\ 2020, \apjl, 900, L44.
\bibitem[Kleisioti et al.(2023)]{kleisiotietal23} Kleisioti, E., Dirkx, D., Rovira-Navarro, M., et al.\ 2023, \aap, 675, A57.
\bibitem[Koppelman \& Helmi(2021)]{mwescape} Koppelman, H.~H. \& Helmi, A.\ 2021, \aap, 649, A136. 
\bibitem[Kreidberg et al.(2019)]{kreidbergetal19} Kreidberg, L., Luger, R., \& Bedell, M.\ 2019, \apjl, 877, L15.
\bibitem[Lagos et al.(2021)]{lagosetal} Lagos, F., Schreiber, M.~R., Zorotovic, M., et al.\ 2021, \mnras, 501, 676.
\bibitem[Lainey et al.(2024)]{laineyetal24} Lainey, V., Rambaux, N., Tobie, G., et al.\ 2024, \nat, 626, 280. 
\bibitem[Lammer et al.(2009)]{lammeretal09} Lammer, H., Bredeh{\"o}ft, J.~H., Coustenis, A., et al.\ 2009, \aapr, 17, 181.
\bibitem[Lammers \& Winn(2024)]{lammers-winn-24} Lammers, C. \& Winn, J.~N.\ 2024, \apjl, 968, 1, L12. 
\bibitem[Levesque et al.(2005)]{levesqueetal05} Levesque, E.~M., Massey, P., Olsen, K.~A.~G., et al.\ 2005, \apj, 628, 973.
\bibitem[Levison et al.(1998)]{levisonetal98} Levison, H.~F., Lissauer, J.~J., \& Duncan, M.~J.\ 1998, \aj, 116, 1998.
\bibitem[Lichtenegger et al.(2010)]{lichtenbergeretal10} Lichtenegger, H.~I.~M., Lammer, H., Grie{\ss}meier, J.-M., et al.\ 2010, \icarus, 210, 1.
\bibitem[Limbach et al.(2021)]{limbachetal21} Limbach, M.~A., Vos, J.~M., Winn, J.~N., et al.\ 2021, \apjl, 918, L25. 
\bibitem[Liu et al.(2013)]{liuetal13} Liu, M.~C., Magnier, E.~A., Deacon, N.~R., et al.\ 2013, \apjl, 777, L20. 
\bibitem[Lovelock(1972)]{lovelockgaia} Lovelock, J.~E.\ 1972, Atmospheric Environment, 6, 579. 
\bibitem[Lucas \& Roche(2000)]{lucasroche00} Lucas, P.~W. \& Roche, P.~F.\ 2000, \mnras, 314, 858. 
\bibitem[Luhman(2014)]{luhman14} Luhman, K.~L.\ 2014, \apjl, 786, L18.
\bibitem[MacDonald(1964)]{macdonald64} MacDonald, G.~J.~F.\ 1964, Reviews of Geophysics and Space Physics, 2, 467. 
\bibitem[Malamud \& Perets(2016)]{malamudperets17a} Malamud, U. \& Perets, H.~B.\ 2016, \apj, 832, 160. 
\bibitem[Malamud \& Perets(2017)]{malamudperets17b} Malamud, U. \& Perets, H.~B.\ 2017, \apj, 849, 8. 
\bibitem[Malmberg et al.(2011)]{malmbergetal11} Malmberg, D., Davies, M.~B., \& Heggie, D.~C.\ 2011, \mnras, 411, 859.
\bibitem[Mao et al.(2021)]{mao-etal} Mao, J., Zhou, P., Simionescu, A., et al.\ 2021, \apjl, 918, L17.
\bibitem[Mardling \& Lin(2002)]{mardling-lin-02} Mardling, R.~A. \& Lin, D.~N.~C.\ 2002, \apj, 573, 2, 829. 
\bibitem[McFadden et al.(2007)]{eurtemp} McFadden, L.-A.~A., Weissman, P.~R., \& Johnson, T.~V.\ 2007, Encyclopedia of the solar system , by McFadden, L.-A. Adams.; Weissman, P. R.; Johnson, T. V. Amsterdam ; Boston : Academic, c2007.
\bibitem[Mestschersky(1893)]{mestschersky} Mestschersky, J.\ 1893, Astronomische Nachrichten, 132, 129. 
\bibitem[Meynet et al.(2015)]{meynetetal15} Meynet, G., Chomienne, V., Ekstr{\"o}m, S., et al.\ 2015, \aap, 575, A60. 
\bibitem[Miret-Roig et al.(2022)]{miretroigetal21} Miret-Roig, N., Bouy, H., Raymond, S.\ N., et al.\ 2022, Nature Astronomy, 6, 89.
\bibitem[Mosqueira \& Estrada(2003a)]{mosqueiraestrada03a} Mosqueira, I. \& Estrada, P.~R.\ 2003, \icarus, 163, 198. 
\bibitem[Mosqueira \& Estrada(2003b)]{mosqueiraestrada03b} Mosqueira, I. \& Estrada, P.~R.\ 2003, \icarus, 163, 232. 
\bibitem[Mr{\'o}z et al.(2018)]{mrozetal18} Mr{\'o}z, P., Ryu, Y.-H., Skowron, J., et al.\ 2018, \aj, 155, 121. 
\bibitem[Mr{\'o}z et al.(2019)]{mrozetal19} Mr{\'o}z, P., Udalski, A., Skowron, J., et al.\ 2019, \apjs, 244, 29. 
\bibitem[Mr{\'o}z et al.(2020)]{mrozetal20} Mr{\'o}z, P., Poleski, R., Gould, A., et al.\ 2020, \apjl, 903, L11. 
\bibitem[Monari et al.(2018)]{mwescape} Monari, G., Famaey, B., Carrillo, I., et al.\ 2018, \aap, 616, L9. 
\bibitem[Murray \& Dermott(1999)]{murraydermott} Murray, C.~D. \& Dermott, S.~F.\ 1999.
\bibitem[Musielak et al.(2005)]{musielaketal05} Musielak, Z.~E., Cuntz, M., Marshall, E.~A., et al.\ 2005, \aap, 434, 355. 
\bibitem[Namouni \& Zhou(2006)]{namouni-zhou} Namouni, F. \& Zhou, J.~L.\ 2006, Celestial Mechanics and Dynamical Astronomy, 95, 245.
\bibitem[Namouni(2005)]{namouni} Namouni, F.\ 2005, \aj, 130, 280.
\bibitem[Nguyen et al.(2022)]{nguyenetal22} Nguyen, C.~T., Costa, G., Girardi, L., et al.\ 2022, \aap, 665, A126. 
\bibitem[Noyola et al.(2014)]{noyolaetal14} Noyola, J.~P., Satyal, S., \& Musielak, Z.~E.\ 2014, \apj, 791, 25. 
\bibitem[Noyola et al.(2016)]{noyolaetal16} Noyola, J.~P., Satyal, S., \& Musielak, Z.~E.\ 2016, \apj, 821, 97. 
\bibitem[Ogihara \& Ida(2012)]{ogiharaida12} Ogihara, M. \& Ida, S.\ 2012, \apj, 753, 60. 
\bibitem[Ogilvie \& Lin(2004)]{ogilvielin04} Ogilvie, G.~I. \& Lin, D.~N.~C.\ 2004, \apj, 610, 477. 
\bibitem[Parriott \& Alcock(1998)]{parriott-alcock} Parriott, J. \& Alcock, C.\ 1998, \apj, 501, 357.
\bibitem[Peale et al.(1979)]{pealeetal79} Peale, S.~J., Cassen, P., \& Reynolds, R.~T.\ 1979, Science, 203, 892. 
\bibitem[Pearson \& McCaughrean(2023)]{pearson-mccaughrean-23} Pearson, S.~G. \& McCaughrean, M.~J.\ 2023, , arXiv:2310.01231. 
\bibitem[Peters-Limbach \& Turner(2013)]{petersturner13} Peters-Limbach, M.~A. \& Turner, E.~L.\ 2013, \apj, 769, 98. 
\bibitem[Porco et al.(2006)]{porcoetal06} Porco, C.~C., Helfenstein, P., Thomas, P.~C., et al.\ 2006, Science, 311, 1393. 
\bibitem[Quanz et al.(2012)]{quanzetal12} Quanz, S.~P., Lafreni{\`e}re, D., Meyer, M.~R., et al.\ 2012, \aap, 541, A133.
\bibitem[Quarles et al.(2020)]{quarlesetal20} Quarles, B., Li, G., \& Rosario-Franco, M.\ 2020, \apjl, 902, L20. 
\bibitem[Rabago \& Steffen(2018)]{rabagosteffen18} Rabago, I. \& Steffen, J.~H.\ 2018, \aas, 231, 148.23
\bibitem[Rasio(1994)]{rasio} Rasio, F.~A.\ 1994, \apjl, 427, L107.
\bibitem[Razbitnaya(1985)]{razbitnaya} Razbitnaya, E.~P.\ 1985, \sovast, 29, 684
\bibitem[Reg{\'a}ly et al.(2022)]{regalyetal22} Reg{\'a}ly, Z., Fr{\"o}hlich, V., \& Vink{\'o}, J.\ 2022, \apj, 941, 121. 
\bibitem[Reg{\'a}ly et al.(2025)]{regalyetal25} Reg{\'a}ly, Z., Fr{\"o}hlich, V., \& Vink{\'o}, J.\ 2025, \apjl, 988, 1, L7. 
\bibitem[Reynolds et al.(1987)]{reynoldsetal87} Reynolds, R.~T., McKay, C.~P., \& Kasting, J.~F.\ 1987, Advances in Space Research, 7, 5, 125. 
\bibitem[Rice et al.(2008)]{riceetal08} Rice, W.~K.~M., Armitage, P.~J., \& Hogg, D.~F.\ 2008, \mnras, 384, 3, 1242.
\bibitem[Roccetti et al.(2023)]{roccettietal23} Roccetti, G., Grassi, T., Ercolano, B., et al.\ 2023, International Journal of Astrobiology, 22, 317. 
\bibitem[Romanova et al.(2023)]{romanovaetal23} Romanova, M.~M., Koldoba, A.~V., Ustyugova, G.~V., et al.\ 2023, \mnras, 523, 2, 2832. 
\bibitem[Romanova et al.(2024)]{romanovaetal24} Romanova, M.~M., Koldoba, A.~V., Ustyugova, G.~V., et al.\ 2024, \mnras, 532, 3, 3509. 
\bibitem[Sajadian \& Sangtarash(2023)]{sajadiansangtarash23} Sajadian, S. \& Sangtarash, P.\ 2023, \mnras, 520, 5613.
\bibitem[Salpeter(1955)]{salpeterxx} Salpeter, E.~E.\ 1955, \apj, 121, 161.
\bibitem[Setiawan et al.(2010)]{setiawan-etal} Setiawan, J., Klement, R.~J., Henning, T., et al.\ 2010, Science, 330, 1642.
\bibitem[Simon et al.(2007)]{simonetal07} Simon, A., Szatm{\'a}ry, K., \& Szab{\'o}, G.~M.\ 2007, \aap, 470, 727.
\bibitem[Simon et al.(2010)]{simonetal10} Simon, A.~E., Szab{\'o}, G.~M., Szatm{\'a}ry, K., et al.\ 2010, \mnras, 406, 2038.
\bibitem[Spencer et al.(2006)]{enctemp} Spencer, J.~R., Pearl, J.~C., Segura, M., et al.\ 2006, Science, 311, 1401.
\bibitem[Stevenson(1999)]{stevenson99} Stevenson, D.~J.\ 1999, \nat, 400, 32.
\bibitem[Strom et al.(2024)]{miranda} Strom, C., Nordheim, T.~A., Patthoff, D.~A., et al.\ 2024, \psj, 5, 226. 
\bibitem[Sumi et al.(2011)]{sumietal11} Sumi, T., Kamiya, K., Bennett, D.~P., et al.\ 2011, \nat, 473, 349.
\bibitem[Sz{\"o}lgy{\'e}n et al.(2022)]{szolgyenetal} Sz{\"o}lgy{\'e}n, {\'A}., MacLeod, M., \& Loeb, A.\ 2022, \mnras, 513, 5465.
\bibitem[Tamayo et al.(2017)]{tamayo-etal-17} Tamayo, D., Rein, H., Petrovich, C., et al.\ 2017, \apjl, 840, 2, L19. 
\bibitem[Teachey \& Kipping(2018)]{teacheykipping18} Teachey, A. \& Kipping, D.~M.\ 2018, Science Advances, 4, eaav1784.
\bibitem[Thomas et al.(2016)]{thomasetal16} Thomas, P.~C., Tajeddine, R., Tiscareno, M.~S., et al.\ 2016, \icarus, 264, 37.
\bibitem[Tjoa et al.(2020)]{tjoaetal20} Tjoa, J.~N.~K.~Y., Mueller, M., \& van der Tak, F.~F.~S.\ 2020, \aap, 636, A50.
\bibitem[Tokadjian \& Piro(2023)]{tokadjianpiro23} Tokadjian, A. \& Piro, A.~L.\ 2023, \aj, 165, 173. 
\bibitem[Veras \& Tout(2012)]{verastout12} Veras, D. \& Tout, C.~A.\ 2012, \mnras, 422, 1648. 
\bibitem[Veras et al.(2011)]{verasetal11} Veras, D., Wyatt, M.~C., Mustill, A.~J., et al.\ 2011, \mnras, 417, 2104.
\bibitem[Veras \& Raymond(2012)]{verasraymond12} Veras, D. \& Raymond, S.~N.\ 2012, \mnras, 421, L117.
\bibitem[Verhulst(1969)]{verhulst} Verhulst, F.\ 1969, \bain, 20, 215
\bibitem[Vink{\'o} et al.(2004)]{vinko04} Vink{\'o}, J., Blake, R.~M., S{\'a}rneczky, K., et al.\ 2004, \aap, 427, 453.
\bibitem[Weldon et al.(2025)]{weldonetal25} Weldon, G.~C., Naoz, S., \& Hansen, B.~M.~S.\ 2025, \apjl, 980, 2, L31. 
\bibitem[Williams et al.(1997)]{williamsetal97} Williams, D.~M., Kasting, J.~F., \& Wade, R.~A.\ 1997, \nat, 385, 234. 
\bibitem[Williams(2013)]{williams13} Williams, D.~M.\ 2013, Astrobiology, 13, 315.
\bibitem[Yeomans(2006)]{eurdensity}Yeomans, D.~K.\ 2006, Planetary Satellite Physical Parameters, NASA JPL
\bibitem[Yoder \& Peale(1981)]{yoderpeale81} Yoder, C.~F. \& Peale, S.~J.\ 1981, \icarus, 47, 1. 
\bibitem[Yoder(1979)]{yoder79} Yoder, C.~F.\ 1979, \nat, 279, 767.
\bibitem[Yoo et al.(2004)]{yoonetal04} Yoo, J., DePoy, D.~L., Gal-Yam, A., et al.\ 2004, \apj, 603, 139. 
\bibitem[Zapatero Osorio et al.(2000)]{zapateroosioroetal00} Zapatero Osorio, M.~R., B{\'e}jar, V.~J.~S., Mart{\'\i}n, E.~L., et al.\ 2000, Science, 290, 103.

\end{thebibliography}
\end{document}